\documentclass[12pt]{JHEP3}
\usepackage{epsfig}
\epsfclipon
\usepackage{multicol}
\usepackage{epsfig,bm}
\usepackage{amssymb,amsmath}
\usepackage{graphicx}

\newcommand{\be}{\begin{equation}}
\newcommand{\ee}{\end{equation}}
\newcommand{\bea}{\begin{eqnarray}}
\newcommand{\beas}{\begin{eqnarray*}}
\newcommand{\eea}{\end{eqnarray}}
\newcommand{\eeas}{\end{eqnarray*}}
\newcommand{\ba}{\begin{array}}
\newcommand{\ea}{\end{array}}

\newcommand{\lb}{\label}

\newcommand{\alp}{\alpha'}

\def\ls{\mathrel{\lower4pt\vbox{\lineskip=0pt\baselineskip=0pt
           \hbox{$<$}\hbox{$\sim$}}}}
\def\gs{\mathrel{\lower4pt\vbox{\lineskip=0pt\baselineskip=0pt
           \hbox{$>$}\hbox{$\sim$}}}}

\newcommand{\N}{\widetilde{N}}

\def\smiley{\hbox{\large$\bigcirc$\hspace{-.80em}%
\raise.2ex\hbox{$\cdot\cdot$}\kern-.61em    
\lower.2ex\hbox{\scriptsize$\smile$}}\ }

\newcommand{\roughly}[1]{\mathrel{\raise.3ex\hbox{$#1$\kern-0.85em
\lower1ex\hbox{$\sim$}}}}

\def\be{\begin{equation}}
\def\beq\begin{equation}
\def\ee{\end{equation}}
\def\bea{\begin{eqnarray}}
\def\eea{\end{eqnarray}}

\def\beq{\begin{equation}}
\def\eeq{\end{equation}}
\def\beqa{\begin{eqnarray}}
\def\eeqa{\end{eqnarray}}

\newcommand{\bmat}{\left(\begin{array}}
\newcommand{\emat}{\end{array}\right)}

\newcommand{\eq}[1]{\mbox{Eq.~(\ref{#1})}}

\title{New developments on embedding inflation in gauge theory and
particle physics}

\author{ A. Mazumdar~$^{1,~2,~3}$\\ $^{1}$~NORDITA,
Blegdamsvej-17, Copenhagen-2100, Denmark \\ $^{2}$~Niels Bohr
Institute, Blegdamsvej-17, Copenhagen-2100, Denmark\\ $^{3}$~Physics
Department, Lancaster University, LA1 4YB, United Kingdom}

\abstract{In this brief review we will discuss how a well motivated
particle theory beyond the eletroweak Standard Model provides
ingredients and conditions for a successful inflation. We will mainly
focus on a low energy supersymmetric Standard Model which provides
plenty of scalars.  In particular, these scalars span a
multidimensional moduli space of {\it gauge invariant} operators which
carry the Standard Model charges. The inflationary predictions which
matches the current observations are robust due to the fact that
inflation occurs within our own gauge sector where the couplings are
well known. We further argue that based on our current understandings
if there exists a {\it string landscape} of multiple vacua, then it is
very natural that the last phase of inflation would be driven by one
of the many supersymmetric Standard Model modulii. Only such a
graceful exit from inflation would provide hot thermal Standard Model
baryons, cold dark matter, conditions for baryogenesis and foremost
the seed density perturbations for the cosmic microwave background
radiation in just {\it one package}.  Furthermore we will also
discuss how some of the ingredients of inflation can be tested already
by the LHC.}

\begin{document}

\section{Introduction}

Primordial inflation as a paradigm has met with glorious successes over
the past $26$ years since its
advent~\cite{Guth,Starobinsky,Linde1,Linde2,Steinhardt} (for a review
see \cite{Lindebook}).  The virtues of inflation lies due to the fact
that an accelerated expansion of the universe can give rise to a
homogeneous and an isotropic universe on very large scales with a flat
spatial geometry. There are very important observational consequences,
such as nearly scale invariant tiny perturbations imparted to the
cosmic microwave background radiation and its spectrum~\cite{Mukhanov}
(for a review see~\cite{Mukhanov-Rev}) has met with an unprecedented
success with the observations based on Cosmic Microwave Background
(CMB) radiation, see the recent data from WMAP~\cite{WMAP3,KKMR}.

It has been known for a while that inflation can be driven by a
dynamical scalar field known as the {\it inflaton}, an order
parameter, which could either be fundamental or
composite. Particularly if the inflaton rolls very slowly on a
sufficiently flat potential (such that the potential energy density
dominates over the kinetic term) then it is possible to show: (1) a
considerably large number of e-foldings, (2) during inflation metric
perturbations are generated with an amplitude: ${\cal P}_{R}\sim
1.91\times 10^{-5}k^{n-1}$, with a spectral tilt lying within:
$0.92\leq n\leq 1.0$ within $2\sigma$ (CMB+SDSS) (3) running of the
spectral tilt is negligible $-0.012 \leq dn_s/d\ln k\leq 0.001$ within
$2\sigma$ (CMB+SDSS for tensor to scalar ratio negligible) and, (4)
the tensor to scalar ratio, $r<0.55$ at $95\%$ c.l. details can be
found in Ref.~\cite{WMAP3}.

However, inspite of all these achievements, there are some fundamental
issues where we are lacking proper understanding, such as:


\begin{itemize}

\item{ What is the origin of inflation ?}

\item{ What is the inflaton ?}

\item{ What are the fundamental interactions of an inflaton ?}

\item{ Where does the inflaton energy goes ?}

\item{ How can we test the inflaton in a laboratory ?}

\end{itemize}

Inspite of many attempts there has been no single good candidate for
an inflaton which comes naturally out of a well motivated theory of
particle physics (for a review on models of inflation,
see~\cite{Lyth-Riotto}). One always relies on scalar fields which are
{\it absolute gauge singlets} possibly residing in some hidden sector
or secluded sector with a small coupling to the SM gauge group. By
definition an {\it absolute gauge singlet} does not carry any charge
what so-ever be the case. Therefore the masses, couplings and
interactions are not generally tied to any fundamental theory or any
symmetry.  Such gauge singlets are used ubiquitously by model
builders to obtain a desired potential and interactions at a free will
in order to explain the current CMB data. In this respect the inflaton
is just a phenomenological answer to inflation and its observed
consequences.  What has been ignored so far is the fact that if
inflation explains the CMB anisotropy then it must also reheat the
universe with the SM degrees of freedom~\footnote{An {\it absolute
gauge singlet} inflaton can couple to all the species and sectors
alike. It does not discriminate between the SM degrees of freedom to
that of any other degrees of freedom. By definition an {\it absolute
gauge singlets} can couple to any other gauge singlets, there is no
symmetry which prohibits doing so. The most popular slow roll
inflationary models such as chaotic inflation~\cite{Linde2}, hybrid
inflation~\cite{Linde-hybrid}, new inflation~\cite{Linde1,Steinhardt},
assisted inflation~\cite{assist1,ninflation}, without slow roll
inflation~\cite{Oscillations}, K-inflation~\cite{K}, etc. all rely on
absolute gauge singlets, see for a review~\cite{Lyth-Riotto}. In this
respect they all fail to explain why the couplings, masses and
interactions are chosen so to explain the CMB data. Moreover one
cannot predict definitely certain issues such as radiative corrections
and stability of the potential, because the couplings and masses are
all put by hand.}. The decay of inflaton should also provide thermal
conditions~\cite{Andy} to create matter-anti-matter asymmetry and
thermal/non-thermal generation of cold dark matter, for a review
see~\cite{MSSM-REV,Kusenko}~\footnote{The Big Bang Nucleosynthesis
(BBN) demands that when the universe was nearly $1$~MeV, most of the
energy density of the universe must lie in the SM relativistic degrees
of freedom~\cite{BBN}. It is therefore pertinent that the universe
must have created a thermal bath of strictly speaking SM degrees of
freedom before the BBN. If there were other degrees of freedom which
were to couple very weakly to the SM then it would provide severe
experimental constraints~\cite{Barger2003}.}.

On the other hand string theory comes up with a plethora of {\it
absolute gauge singlet} modulii, which mainly arise in the
gravitational sector upon various
compactifications~\cite{Kachru}. There are numerous attempts to embed
inflation within string theory, for a review see~\cite{Kallosh}, see
also~\cite{KKLT,KKLMMT}. However they all tend to concentrate on
explaining the CMB data, but with very little to do with an aftermath
of inflation~\cite{Myers}, such as how the inflaton couples to the SM
degrees of freedom ?, how the dark matter is created ?, what are the
relativistic degrees of freedom after the end of inflation ?, etc. The
uncertainties prevail mainly due to two facts: (a) It is not clear how
to construct SM like gauge group. There are various constructions so
far, but all suffer through various problems such as quantum
instability, moduli de-stabilization, extra U(1)'s,
etc.~\cite{Lust}. (b) Inflation again generically driven by absolute
gauge singlets, such as the closed string modulii in the case of a
race-track inflation~\cite{Burgess}, or open string
moduli~\cite{Dvali,Panda}, for a recent discussion on open string
moduli inflation in a warped throat, see~\cite{McAllister}. In the
case of an open string moduli the inflaton need not be completely an
absolute gauge singlet, but so far a realistic model of
brane-anti-brane inflation lacks the SM embedding~\cite{Horace}.

In all these cases the inflaton candidate can couple to closed string
degrees of freedom and the open string degrees of freedom
alike. Although there will be some preference to decay channels for an
open string moduli as an inflaton, it can couple to an appropriate
gauge theory with appropriate gauge couplings~\cite{Horace}. However
there is no obvious construction which connects open string moduli as
an inflaton to that of the SM sector.  Moreover because of a
hierarchy between the four dimensional Planck scale, string scale and
the compactification scale, the $6$ dimensional compactified volume is
so large that a kinematical phase factor argument allows the inflaton
energy density to get lost in the bulk of the volume which contains
mainly the closed string degrees of freedom~\cite{Panda,Cline}.
Therefore, there is no preference why the inflaton would decay
primarily into the SM baryons.  One would have to make additional
assumptions regarding various sectors the inflaton couples to or where
exactly inflation occurs~\footnote{In a warped throat geometry,
motivated by stabilizing the dilaton, complex structure modulii and
the k\"ahler modulii, the inflaton sector and the SM sector are kept
in distinctive throats. The SM throat typically addresses the
hierarchy issue while the inflationary throat is situated relatively
at higher energies. The two throats talk to each other through closed
string sector. There are many issues which plague this geometry, such
as excess Kaluza Klein graviton production~\cite{Koffman-Yi,Myers},
and transferring the energy density from closed string sector to the
SM throat~\cite{Myers}. There are various subtleties as during
inflation the SM modulii are displaced away from their minimum which
brings in various cosmological uncertainties which might as well
hamper baryogenesis and BBN~\cite{Myers}.}. The string motivated
models have thus failed to explain the aftermath of inflation,
particularly why the universe is filled with the SM baryons at the
time of BBN. More work is required to understand this issue, until
then one could at best treat stringy inflation as a nice playground to
test various ideas concerning CMB data alone.

There are also some attempts to understand the CMB data without
invoking inflation such as in the case of a bouncing
cosmology~\cite{Ekpyrotic,Siegel,Brandenberger}. Some of the key ideas
are stringy in origin (based on t-duality, see~\cite{BV}), especially
in the context of string gas cosmology of Brandenberger and
Vafa~\cite{BV}. However, there are various caveats towards a full
understanding of a thermodynamics of a gas of strings in a
cosmological background~\cite{Frey}. There are other attempts to
obtain inflation such as in the case of a gas of
D-branes~\cite{Easson}, or string moduli in conjunction with string
gas~\cite{Biswas}, or in the case of Hagedorn strings during a
bounce~\cite{BBSM}.  Moreover they all fail to address one crucial
issue: how to obtain the SM degrees of freedom at the end of the day?

An an important message from the above discussions is following:\\
{\it If any model of inflation or any alternatives of inflation
provides the right CMB predictions then there must be a way to excite
the SM baryons with a precision which can match the BBN data without
making further assumptions.}\\ In order to facilitate this we need a
guiding principle, some kind of symmetry argument, which will help us
in constructing a successful model of inflation from a bottom-up
approach.

Very recently an interesting construction of inflation has been
suggested within a minimal extension of the SM such as in the case of
Minimal Supersymmetric Standard Model
(MSSM)~\cite{AEGM,AEGJM,AKM,AJM,ADM,AFM,AM} (for a review on MSSM,
see~\cite{Kane,Haber} and its embedding in supergravity (SUGRA),
see~\cite{sugra,sugra1,Nilles}, for cosmological aspects of MSSM,
see~\cite{MSSM-REV,Kusenko}. Within MSSM there are many scalars, which
span into a moduli space of {\it gauge invariant} F-and D-flat
directions, which carry SM charges usually baryon and/or lepton
number. These modulii have an {\it enhanced symmetry point} near the
origin (at a VEV defined by zero). Away from the origin these modulii
break wholly or partly the SM gauge symmetry depending on the flat
direction. But such a spontaneous breaking of charge and color in the
early universe is not considered to be dangerous, provided they all
settle down to their minimum before the electroweak phase transition.

In the first paper~\cite{AEGM}, it was pointed out that the MSSM
provides all the necessary key ingredients for a successful
inflation. Out of nearly $300$ flat directions~\cite{MSSM-REV}, there
are only $2$ directions which can support inflation with a graceful
exit. Other directions do not have a graceful exit~\cite{AEGM,AEGJM},
nor do they have a slow roll phase of inflation. This is a remarkable
result which puts forward the two inflaton candidates as $LLe$ (which
carries the lepton number) and, $udd$ (which carries the baryon
number). The $L$ corresponds to the slepton (SUSY partner of SM
lepton) doublet, while $e$ corresponds to the right handed selectron
(SUSY partner of electron). The $u,~d,~d$ correspond to the right
handed squarks (SUSY partner of quarks). These two directions can
support {\it eternal} and {\it slow roll} inflation with a right
amplitude of temperature anisotropy ${\cal P}_{R}\sim \delta_{H}\sim
1.91\times 10^{-5}$, an observable tilt in the power spectrum with a
range $0.92 \leq n_s\leq 1.0$, where the lower limit is saturated for
a {\it saddle point} inflation, the upper limit corresponds to having
a slow roll inflation away from the saddle point~\cite{AEGM,AEGJM}.
The running of the tilt in the spectrum is negligible and the tensor
to scalar ratio is also observationally negligible. Moreover there is
no production of cosmic strings, or non-MSSM degrees of
freedom~\cite{AEGM,AEGJM}, and there is no large
non-Gaussianity~\cite{Instant-NG}.

Furthermore the inflaton reheating is very well understood as the
inflatons carry the SM charges, they naturally decay only to the SM
degrees of freedom, besides creating the lightest supersymmetric
particle (LSP) as a candidate for the cold dark
matter~\cite{AEGM,AEGJM,ADM}. With an R-parity LSP's are absolutely
stable and can be a SUSY cold dark matter
candidate~\cite{Bertone,Ellis}. In Ref.~\cite{ADM} we studied the
parameter space of inflation and the neutralino type dark matter
produced thermally. Remarkably we found an interesting overlap between
inflationary parameters within $m_0,~m_{1/2}$ (soft masses and gaugino
masses respectively) plane in (m-SUGRA) setup. This provides another
hint that if the desert is filled by MSSM from the electroweak scale
to the grand unified scale, then the parameters of an MSSM inflaton
will provide not only a successful inflation which matches the current
observational data, but the same parameter space is also responsible
for generating the observed abundance of thermally created cold dark
matter~\cite{ADM}.

Both the inflatons, $LLe,~udd$, masses are tied with a low energy SUSY
required to address the hierarchy between the Planck and the
electroweak scale. The mass range $\sim {\cal O}(\rm TeV)$ is adequate
to explain the CMB anisotropy. An arbitrary increase in the scalar
masses will ruin the CMB predictions~\cite{AEGM,AEGJM}. Not only this,
we would also be able to put constraints on the slepton and squark
masses from the LHC~\cite{ADM}. This will further constrain the MSSM
inflation model to an unprecedented level. In this regard both $LLe$ and
$udd$ are testable through CMB and the LHC. This is the most exciting
expectation that for the first time the LHC will be able to rule out a
model of inflation completely. For instance if the slepton and squark
masses are beyond $\gg 10$~TeV, then this model of inflation is ruled
out completely. For such a mass range neither $LLe$ nor $udd$ can
generate the right amplitude of density
perturbations~\cite{AEGM,AEGJM}.

Furthermore this is the only known model of inflation where the ${\cal
N}_{\rm COBE}$ (the last number of e-foldings required to explain the
CMB data) is determined fully from the fundamental interactions. The
${\cal N}_{\rm COBE}$ is primarily determined by thermal history of
the universe. Fortunately enough, in MSSM based inflationary model,
the reheating and thermalization temperature can be estimated rather
accurately due to known SM couplings~\cite{AEGM,AEGJM,AKM,AJM}.

Another important issue is that within MSSM inflation the quantum
stability of the inflationary potential can be analyzed correctly as
the couplings of the inflaton are known~\cite{AEGJM}. Since inflation
is driven near the saddle point, it becomes an important question to
ask how stable the potential is under large radiative corrections?  To
our surprise what we found is that an existence of a saddle point is
not ruined by radiative corrections, although the point of inflection
does shift towards higher vacuum expectation values (VEVs). The
predictions for CMB also does not modify at all within the current
uncertainties.  The model predictions are also robust as there is no
supergravity (SUGRA) corrections which can spoil the {\it cosmological
flatness} of the potential. Although one would expect that SUGRA to
play an important role, but all such corrections are absorbed when the
choice of a saddle point is made where slow roll inflation occurs. In
this respect one can take a view that a saddle point inflation can
address the SUGRA-eta problem~\cite{AEGJM}.  Of course SUSY is the key
ingredient which helps maintaining the flatness of the potential, in
this respect SUSY along with gauge symmetry not only guides us towards
a more holistic model of inflation, but also with sharp predictions
which can be tested at the LHC and in future CMB experiments. Note
that the model does not generate observationally significant
stochastic gravitational wave background radiation or the cosmic
strings.

An interesting twist comes from string theory alone. As it is fairly
well established by now that there exists a string theory landscape
with plethora of vacua, for a review~\cite{Kachru}.  Therefore, in
this landscape, only way our patch of the observable universe evolves
to its present state regardless of how it began, if we only secure
that the last stage of inflation were driven within an MSSM vacuum. In
any case the MSSM fields (and couplings) must be in the low energy
spectrum. At some stage the MSSM sector should take over. False vacuum
inflation in MSSM then makes any previous stage in the history of the
universe oblivious. However the observable part of the universe can
still emerge since false vacuum inflation also sets the stage for a
last bout of successful inflation (also happening in the MSSM sector).

In the following sections we will discuss various virtues of MSSM
inflation, in section 2, we discuss general principles behind a
successful inflation, in section 3, we discuss MSSM and introduce a
concept behind F-and D- flat directions. In section 4, we describe
MSSM inflation. In section 5, we discuss radiative and SUGRA
corrections. In section 6, we discuss reheating and thermalization. In
section 7, we discuss thermal neutralino cold dark matter creation.
In section 8, our effort is to embed MSSM inflation with grand unified
theories with an inclusion of the Majorana neutrinos.  In section 9,
we discuss how a small Yukawa coupling required to explain the
neutrino masses could also maintain the flatness of the MSSM
inflationary potential. In the same section we describe MSSM inflation
within gauge mediated SUSY breaking scenarios. In the last section 10,
we discuss initial conditions for an MSSM inflation, we speculate the
role of string landscape in order to connect high and low scale
inflations. We also point out a graceful exit from a string landscape
which can happen via a slow roll MSSM inflation.


\section{What should be the inflation properties?}

In a minimal model, the inflaton is the only source of density
perturbations and also responsible for a successful reheating. This is
also true in the case of a curvaton~\cite{Curvaton}. Where the success
of curvaton lies within a successful reheating of the SM degrees of
freedom during the curvaton dominance. It is possible to find an MSSM
curvaton candidate which has all the properties to create an universe
with the right relativistic degrees of freedom, cold dark matter and
the seed for the primordial density perturbations~\cite{We}. There are
also variants of curvaton in the context of inhomogeneous reheating
and generating density perturbations~\cite{Inhomogeneous}. In either
cases it is important to realize that a successful reheating ensures a
successful CMB predictions, as an example in the case of an MSSM
curvaton, recently we have singled out an unique MSSM flat direction
which would create the right CMB predictions and the SM baryons along
with neutralino type cold dark matter~\cite{AEJM}.

The dream has been to embed the model of inflation in particle physics
naturally. The first attempts were made by Guth~\cite{Guth}, where
inflation would occur in a false vacuum of the Higgs of the Grand
Unified Theory (GUT), fell into trouble, because the end of inflation
happened due to first order phase transition. The bubble of true
vacuum remains cold in the sea of false vacuum and there is no way to
generate thermal entropy other than colliding the bubbles ( the
potential energy is stored in the bubble walls ). There were attempts
to address this issue, but all such attempts were not so attractive,
as they would all involve fields such as absolute gauge singlets from
a hidden sector, see for a review~\cite{Lyth-Riotto}.

A successful reheating must ensure that:

\begin{itemize}

\item{SM Baryons are excited:\\The SM baryons are excited dominantly
as light degrees of freedom to ensure a successful BBN~\cite{BBN}.
Note that BBN puts stringent constraints on hidden {\it light} degrees
of freedom~\cite{Barger-2003}. At best one could accommodate one light
neutrino or any relativistic species~\cite{BBN}. This suggests that
the hidden sector particles must be heavier than few $\sim$~ TeV so
that they kinetically and thermally decouple from the rest of the
thermal bath before they could decay through Planck suppressed
interactions.}

\item{Cold dark matter is created:\\ A successful reheating must
create conditions for generating thermally/non-thermally cold dark
matter essential for the structure formation~\cite{WMAP3}. }

\end{itemize}

These two conditions do not necessarily bar the inflaton to be a part
of a hidden sector. However if our main aim is to seek a model of
particle physics then the hidden sector serves very little progress in
our quest. Note that there is no underlying symmetry which helps
understanding the masses of the hidden sector fields and their
couplings to the SM degrees of freedom.

If any model of inflation seeks to be successful, then it must possess:

\begin{itemize}

\item{Credibility:\\ The model parameters, such as mass, couplings are
not chosen ad-hoc to match the CMB observations, rather they should
arise {\it naturally} from an underlying theory.}

\item{Stability:\\ As it is well known a slow roll inflation needs a
flat potential. By construction the inflaton energy density couples to
gravity and the inflaton couples to at least the SM degrees of
freedom, therefore, one needs to ensure that the background geometry,
quantum corrections, supergravity effects do not spoil the flatness of
the potential. }

\item{Testability:\\ It would be desirable to test the micro physical
ingredients of inflation in a terrestrial laboratory. It is fair to
say that we have indeed tested several ideas of inflation in CMB
physics, but in order to really seek the true origin of inflation one
must do more than that. }

\end{itemize}

Based on the above issues, it is arguably simpler if inflation were
driven solely by the SM particles. However SM spectra is full of
fermions, only scalar is the SM Higgs but with a relatively small VEV
to match the observational data~\footnote{Could SM Higgs be the
inflaton ?\\ The Higgs searches at LEP has pushed the Higgs mass above
$114$~GeV~\cite{LEP}, this means that the electroweak phase transition
via the SM Higgs is of second order or perhaps crossover in
nature. Inflation could potentially work out if the Higgs field rolls
very slowly, however, the Higgs VEV is sufficiently low enough to
generate the scalar density perturbations to match the observed
temperature anisotropy.}.

This would eventually push the inflaton candidature to the physics
beyond the SM. However this is the frontier where we lack our grounds
on the experimental front, hence we are forced to the speculations.
This is perhaps one of the main reasons for introducing {\it absolute
gauge singlets} as the inflaton, or modeling the inflaton in a {\it
hidden sector}. However, here as we have reiterated, we wish to have a
concrete model of inflation which has a better predictions from the
CMB observations, successful reheating and of course the model should
be testable in future collider.


\section{Supersymmetry as a tool}

Supersymmetry (SUSY) has many virtues, foremost, it is the most
attractive scenario to address the hierarchy between the Planck and the
electroweak scale. One particular advantage of SUSY is that the
quantum corrections due to bosonic and fermionic loops cancel exactly
in a SUSY limit, rendering the stability of masses, couplings and the
scalar potentials. In order to address the hierarchy problem the SUSY
must be broken spontaneously. At low energies, within MSSM, the SUSY
breaking effects are captured by {\it soft} parameters, i.e. mass,
trilinear couplings, etc.  In low energy SUSY breaking scenarios, the
scalar mass is $\sim {\cal O}(\rm TeV)$. Further note that SUSY allows
many scalar fields (corresponding to every quarks and leptons within
MSSM), therefore, it is interesting to ask what is the role of {\it
sfermions} in cosmology?

In {\it softly} broken SUSY, the quantum corrections give rise to
Logarithmic running to masses and couplings. This ensures that the
scalar potential is at least Logarithmically stable under quantum
corrections. In this respect the SUSY inspired inflationary potentials
are at least stable under quantum corrections. This leads to
satisfying one of the cornerstones of a successful inflationary model
building.


\subsection{MSSM and its potential}

Let us remind the reader that the matter fields of MSSM are chiral
superfields $\Phi=\phi+\sqrt{2}\theta\bar\psi+\theta\bar\theta F$, which
describe a scalar $\phi$, a fermion $\psi$ and a scalar auxiliary
field F. In addition to the  usual quark and lepton superfields,
MSSM has two Higgs fields, $H_u$ and $H_d$. Two Higgses are needed
because $H^\dagger$, which in the Standard Model gives masses
to the $u$-quarks,  is forbidden in the superpotential.

The superpotential for the MSSM is given by~\cite{Nilles}
\begin{equation}
\label{mssm}
W_{MSSM}=\lambda_uQH_u u+\lambda_dQH_d d+\lambda_eLH_d e~
+\mu H_uH_d\,,
\end{equation}
where $H_{u}, H_{d}, Q, L,~u,~d,~ e$ in Eq.~(\ref{mssm}) are chiral
superfields, and the dimensionless Yukawa couplings $\lambda_{u},
\lambda_{d}, \lambda_{e}$ are $3\times 3$ matrices in the family
space. We have suppressed the gauge and family indices. Unbarred
fields are $SU(2)$ doublets, barred fields $SU(2)$ singlets. The last
term is the $\mu$ term, which is a supersymmetric version of the SM
Higgs boson mass.  Terms proportional to $H_{u}^{\ast}H_{u}$ or
$H^{\ast}_{d}H_{d}$ are forbidden in the superpotential, since
$W_{MSSM}$ must be analytic in the chiral fields. $H_{u}$ and $H_{d}$
are required not only because they give masses to all the quarks and
leptons, but also for the cancellation of gauge anomalies. The Yukawa
matrices determine the masses and CKM mixing angles of the ordinary
quarks and leptons through the neutral components of
$H_{u}=(H^{+}_{u},H^{0}_{u})$ and $H_{d}=(H^{0}_{d},H^{-}_{d})$.  Since
the top quark, bottom quark and tau lepton are the heaviest fermions
in the SM, we assume that only the $(3,3)$ element of the matrices
$\lambda_{u}, \lambda_{d}, \lambda_{e}$ are important. In this limit
only the third family and the Higgs fields contribute to the MSSM
superpotential.

The SUSY scalar potential $V$ is the sum of the F- and D-terms and reads
\begin{equation}
\label{fplusd}
V= \sum_i |F_i|^2+\frac 12 \sum_a g_a^2D^aD^a
\end{equation}
where
\begin{equation}
F_i\equiv {\partial W_{MSSM}\over \partial \phi_i},~~D^a=\phi^\dagger T^a
\phi~.
\label{fddefs}
\end{equation}
Here we have assumed that $\phi_i$ transforms under a gauge group
$G$ with the generators of the Lie algebra given by
$T^{a}$.

\subsection{F-and D-flat directions}

For a general supersymmetric model with $N$ chiral superfields
$X_{i}$, it is possible to find out the directions where the potential
Eq.~\ref{fplusd} vanishes identically by solving simultaneously
\begin{equation}
\label{fflatdflat}
D^{a}\equiv X^{\dagger}T^{a}X=0\,, \quad \quad
F_{X_{i}}\equiv \frac{\partial W}{\partial X_{i}}=0\,.
\end{equation}
Field configurations obeying Eq.~(\ref{fflatdflat}) are called respectively
D-flat and F-flat.

D-flat directions are parameterized by gauge invariant monomials of
the chiral superfields. A powerful tool for finding the flat
directions has been developed in \cite{dine96,gherghetta96}, where the
correspondence between gauge invariance and flat directions has been
employed. The configuration space of the scalar fields of the MSSM
contains $49$ complex dimensions ($18$ for $Q_{i}$, $9$ each for $\bar
u_{i}$ and $\bar d_{i}$, $6$ for $L_{i}$, $3$ for $\bar e_{i}$, and
$2$ each for $H_{u}$ and $H_{d}$), out of which there are $12$ real
D-term constraints ($8$ for $SU(3)_{C}$, $3$ for $SU(2)_{L}$, and $1$
for $U(1)_{Y}$), which leaves a total of $37$ complex dimensions
\cite{dine96,gherghetta96}. The trick is to construct gauge invariant
monomials forming $SU(3)_{C}$ singlets and then using them as building
blocks to generate $SU(3)_{C}\times SU(2)_{L}$, and subsequently the
whole $SU(3)_{C}\times SU(2)_{L}\times U(1)_{Y}$ invariant polynomials
\cite{dine96,gherghetta96}. However these invariant monomials give
only the D-flat directions. For F-flat directions, one must solve
explicitly the constraint equations $F_{X_{i}}=0$.

A single flat direction necessarily carries a global $U(1)$ quantum
number, which corresponds to an invariance of the effective Lagrangian
for the order parameter $\phi$ under phase rotation $\phi\to
e^{i\theta}\phi$.  In the MSSM the global $U(1)$ symmetry is
$B-L$. For example, the $LH_u$-direction (see below) has $B-L=-1$.

A flat direction can be represented by a composite gauge invariant
operator, $X_m$, formed from the product of $k$ chiral superfields
$\Phi_i$ making up the flat direction: $X_m=\Phi_1\Phi_2\cdots
\Phi_m$.  The scalar component of the superfield $X_m$ is related to
the order parameter $\phi$ through $X_m=c\phi^m$. For a flat direction
represented by polynomial the description is much more involved,
see~\cite{Jokinen}.

\begin{table}
\vspace{3mm}
\begin{tabular}{|c|c|c|c|}
\hline & $B-L$ & & $B-L$ \\ \hline
$H_{u}H_{d}$ & 0 & $LH_{u}$ &-1\\
$\bar u\bar d\bar d$ & -1 & $QL\bar d$ & -1\\
$LL\bar e$ & -1 & $QQ\bar u\bar d$ & 0\\
$QQQL$ & 0 & $QL\bar u\bar e$ & 0\\
$\bar u\bar u\bar d\bar e $ & 0 & $QQQQ\bar u$ & 1\\
$QQ\bar u\bar u\bar e$ & 1 & $LL\bar d\bar d\bar d$ & -3\\
$\bar u\bar u\bar u\bar e\bar e$ & 1 & $QLQL\bar d\bar d$ & -2\\
$QQLL\bar d\bar d$ & -2 & $\bar u\bar u\bar d\bar d\bar d\bar d$ & -2\\
$QQQQ\bar dLL$ & -1 & $QLQLQL\bar e$ & -1\\
$QL\bar u QQ\bar d\bar d$ & -1 &
$\bar u\bar u\bar u\bar d\bar d\bar d\bar e$ & -1\\
\hline
\end{tabular}
\caption{\label{table1}
{\bf Renormalizable F and D flat directions in the MSSM }}
\end{table}

\subsubsection{An example of F-and D-flat direction}

The flat directions in the MSSM are tabulated in Table 1.
An example of a D-and F-flat direction is provided by
\begin{equation}
\label{example}
H_u=\frac1{\sqrt{2}}\left(\begin{array}{l}0\\ \phi\end{array}\right),~
L=\frac1{\sqrt{2}}\left(\begin{array}{l}\phi\\ 0\end{array}\right)~,
\end{equation}
where $\phi$ is a complex field parameterizing the flat direction,
or the order parameter, or the AD field. All the other fields are
set to zero. In terms of the composite gauge invariant operators,
we would write $X_m=LH_{u}~(m=2)$.

From \eq{example} one clearly obtains $ F_{H_u}^*=\lambda_uQ\bar u
+\mu H_d=F_{L}^*=\lambda_dH_d\bar e\equiv 0$ for all $\phi$. However
there exists a non-zero F-component given by $F^*_{H_d}=\mu
H_u$. Since $\mu$ can not be much larger than the electroweak scale
$M_W\sim {\cal O}(1)$~TeV, this contribution is of the same order as
the soft supersymmetry breaking masses, which are going to lift the
degeneracy. Therefore, following \cite{dine96}, one may nevertheless
consider $LH_u$ to correspond to a F-flat direction.

The relevant D-terms read
\begin{equation}
\label{Dterm0}
D^a_{SU(2)}=H_u^\dagger\tau_3H_u+L^\dagger\tau_3L=\frac12\vert\phi\vert^2
-\frac12\vert\phi\vert^2\equiv 0\,.
\end{equation}
Therefore the $LH_u$ direction is also D-flat.

The only other direction involving the Higgs fields and thus soft
terms of the order of $\mu$ is $H_uH_d$. The rest are purely leptonic,
such as $LL\bar e$, or baryonic, such as $\bar u\bar d\bar d$, or
mixtures of leptons and baryons, such as $QL\bar d$.

\subsubsection{Lifting by non-renormalizable operators}

Non-renormalizable superpotential terms in the MSSM can be viewed as
effective terms that arise after one integrates out fields with very
large mass scales appearing in a more fundamental (say, string)
theory.  Here we do not concern ourselves with the possible
restrictions on the effective terms due to discrete symmetries present
in the fundamental theory, but assume that all operators consistent
with symmetries may arise. Thus in terms of the invariant operators
$X_m$, one can have terms of the type \cite{dine95,dine96}
\begin{equation}
\label{Xton}
W=\frac{h}{d M^{d-3}} X^{k}_{m}=\frac{h}{d M^{d-3}}\phi^d\,,
\end{equation}
where the dimensionality of the effective scalar operator $d=mk$, and
$h$ is a coupling constant which could be complex with $|h|\sim {\cal
O}(1)$. Here $M$ is some large mass, typically of the order of the
Planck mass or the string scale (in the heterotic case $M \sim
M_{GUT}$). The lowest value of $k$ is $1$ or $2$, depending on whether
the flat direction is even or odd under $R$-parity.

A second type of term lifting the flat direction would be of the form
\cite{dine95,dine96}
\begin{equation}
\label{2ndtype}
W={h^{\prime}\over M^{d-3}}\psi\phi^{d-1}~\,,
\end{equation}
where $\psi$ is not contained in $X_m$. The superpotential term
\eq{2ndtype} spoils F-flatness through $F_\psi \neq 0$. An example
is provided by the direction $\bar u_1\bar u_2\bar u_3\bar e_1\bar e_2$,
which is lifted by the non-renormalizable term
$W=(h'/M)\bar u_1\bar u_2\bar d_2\bar e_1$. This superpotential term
gives a non-zero contribution
$F_{\bar d_2}^*=(h'/M)\bar u_1\bar u_2\bar e_1\sim (h^{\prime}/M)\phi^3$
along the flat direction.

Assuming minimal kinetic terms, both types discussed above in
Eqs.~(\ref{Xton},\ref{2ndtype}) yield a generic non-renormalizable
potential contribution that can be written as 
\be
\label{nrpot}
V(\phi)={\vert\lambda\vert^2\over M^{2d-6}}(\phi^*\phi)^{d-1}~,
\ee
where we have defined the coupling $|\lambda|^2\equiv |h|^2+|h'|^2$.
By virtue of an accidental $R$-symmetry under which $\phi$ has a charge
$R=2/d$, the potential \eq{nrpot} conserves the $U(1)$ symmetry carried
by the flat direction, in spite of the fact that at the superpotential
level it is violated, see Eqs.~(\ref{Xton},\ref{2ndtype}).
The symmetry can be violated if there are multiple flat directions,
or by higher order operator contributions. However it turns out
\cite{dine96} that the $B-L$ violating terms are always subdominant.
This is of importance for baryogenesis considerations, where the necessary
$B-L$ violation should therefore arise from other sources, e.g. such
as soft supersymmetry breaking terms.


\section{Gauge Invariant Inflaton}

\begin{figure}
\vspace*{-0.0cm}
\begin{center}
\epsfig{figure=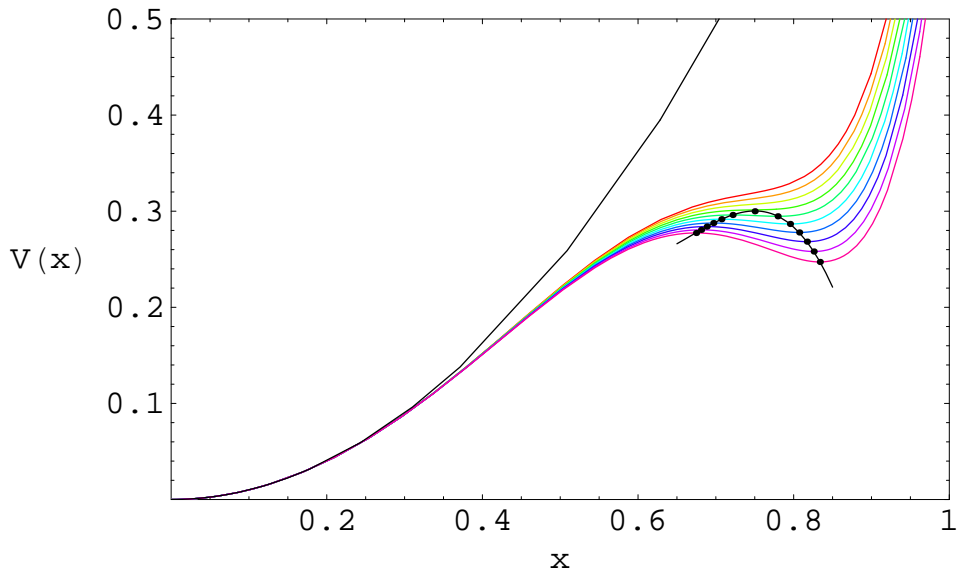,width=.84\textwidth,clip=}
\vspace*{-0.0cm}
\end{center}
\caption{ The colored curves depict the full potential, where
$V(x)\equiv V(\phi)/(0.5~m_{\phi}^2 M_{\rm P}^2(m_{\phi}/M_{\rm
P})^{1/2})$, and $ x\equiv (\lambda_n M_{\rm P}/m_{\phi})^{1/4}
(\phi/M_{\rm P})$. The black curve is the potential arising from the
soft SUSY breaking mass term. The black dots on the colored potentials
illustrate the gradual transition from minimum to the saddle point and
to the maximum.}
\label{fig0-pot}
\end{figure}

Let us recapitulate the main features of MSSM flat direction
inflation~\cite{AEGM,AEGJM,ADM}. The framework is solely based on MSSM
together with gravity, so consistency dictates that all
non-renormalizable terms allowed by gauge symmetry and supersymmetry
should be included below the cut-off scale, which we take to be the
Planck scale. The superpotential term which lifts the $F$-flatness is
given by:
\beq \label{supot}
W_{non} = \sum_{n>3}{\lambda_n \over n}{\Phi^n \over M^{n-3}}\,,
\eeq
where $\Phi$ is a {\it gauge invariant} superfield which contains the
flat direction.  Within MSSM all the flat directions are lifted by
non-renormalizable operators with $4\le n\le 9$~\cite{gherghetta96},
where $n$ depends on the flat direction. We expect that quantum
gravity effects yield $M=M_{\rm P}=2.4\times 10^{18}$~GeV and
$\lambda_n\sim {\cal O}(1)$~\cite{dine95,dine96}. Note however that
our results will be valid for any values of $\lambda_n$, because
rescaling $\lambda_n$ simply shifts the VEV of the flat direction.
Let us focus on the lowest order superpotential term in
Eq.~(\ref{supot}) which lifts the flat direction. Soft SUSY breaking
induces a mass term for $\phi$ and an $A$-term so that the scalar
potential along the flat direction reads
\beq \label{scpot}
V = {1\over2} m^2_\phi\,\phi^2 + A\cos(n \theta  + \theta_A)
{\lambda_{n}\phi^n \over n\,M^{n-3}_{\rm P}} + \lambda^2_n
{{\phi}^{2(n-1)} \over M^{2(n-3)}_{\rm P}}\,,
\eeq
Here $\phi$ and $\theta$ denote respectively the radial and the
angular coordinates of the complex scalar field
$\Phi=\phi\,\exp[i\theta]$, while $\theta_A$ is the phase of the
$A$-term (thus $A$ is a positive quantity with dimension of mass).
Note that the first and third terms in Eq.~(\ref{scpot}) are positive
definite, while the $A$-term leads to a negative contribution along
the directions whenever $\cos(n \theta + \theta_A) < 0$, see
\ref{fig0-pot}~\footnote{The importance of the A-term was first
highlighted in a successful MSSM curvaton model~\cite{AEJM}, where
again the curvaton carries the SM charges. By implementing so it also
leads to a successful CMB prediction, SM reheating and a detectable
signature at the LHC.}. There are other attempts to embed inflation
within a gauge theory~\cite{Others-G}, they however do not explain SM
reheating, besides they also have some caveats related to Planckian
VEVs.


\subsection{Inflation near the Saddle Point}

The maximum impact from the $A$-term is obtained when $\cos(n \theta +
\theta_A) = -1$ (which occurs for $n$ values of $\theta$).  Along
these directions $V$ has a secondary minimum at $\phi = \phi_0 \sim
\left(m_\phi M^{n-3}_{\rm P}\right)^{1/n-2} \ll M_{\rm P}$ (the global
minimum is at $\phi=0$), provided that
\beq \label{extrem}
A^2 \geq 8 (n-1) m^2_{\phi}\,.
\eeq
At this minimum the curvature of the potential is positive both along
the radial and angular directions. If the $A$ is too large, the
secondary minimum will be deeper than the one in the origin, and hence
becomes the true minimum. However, this is {\it phenomenologically}
unacceptable as such a minimum will break charge and/or
color~\cite{dine95,dine96}.  With a total potential: $V \sim
m_\phi^2\phi_0^2 \sim m_{\phi}^2\left(m_{\phi} M^{n-3}_{\rm
P}\right)^{2/(n-2)}$.

As discussed in ~\cite{AEGM}, if the local minimum is too steep, the
field will become trapped there with an ensuing inflation that has no
graceful exit like in the old inflation scenario~\cite{Guth}.  On the
other hand in an opposite limit, with a point of inflection, a single
flat direction cannot support inflation~\cite{Jokinen1}, one would
require an assisted inflation with the help of many flat
directions~\cite{assist1}..

However, in the gravity mediated SUSY breaking case, the $A$-term and
the soft SUSY breaking mass terms are expected to be the same order of
magnitude as the gravitino mass, i.e.
\begin{eqnarray}
\label{condi-0}
m_{\phi}\sim A \sim m_{3/2}\sim {\cal O}(1)~{\rm TeV}\,.
\end{eqnarray}
Therefore, as pointed out in~\cite{AEGM}, in the gravity mediated SUSY
breaking it is possible that the potential barrier actually disappears
and the inequality in Eq.~(\ref{extrem}) is saturated so that $A$ and
$m_\phi$ are related by
\beq
\label{cond}
A^2 = 8 (n-1) m^2_\phi\,.
\eeq
If the above condition is satisfied then both the first and second
derivatives of $V$ vanish at $\phi_0$, ~i.e. $V^{\prime}(\phi_0)=0,~
V^{\prime\prime}(\phi_0)=0$.  As the result, if initially $\phi\sim
\phi_0$, a slow roll phase of inflation is driven by the third
derivative of the potential.

Note that this behavior does not seem possible for other SUSY breaking
scenarios such as the gauge mediated breaking~\cite{GMSB} or split
SUSY~\cite{SPLIT}. In split SUSY the $A$-term is protected by an
$R$-symmetry, which also keeps the gauginos light while the sfermions
are quite heavy~\cite{SPLIT}~\footnote{In the gauge mediated case
there is an inherent mismatch between $A$ and $m_{\phi}$, except at
very large field values where Eq.~(\ref{condi-0}) can be satisfied.
However there exists an unique possibility of a saddle point inflation
within gauge mediated case which we will discuss in Section
9.2~\cite{AJM}.}.


\subsection{Slow roll}

The potential near the saddle point Eq. (\ref{cond}) is very flat
along the {\it real direction} but not along the {\it imaginary
direction}. Along the imaginary direction the curvature is determined
by $m_{\phi}$.  Around $\phi_0$ the field lies in a plateau with a
potential energy
\beq \label{potential}
V(\phi_0) = {(n-2)^2\over2n(n-1)}\,m^2_\phi \phi_0^2
\eeq
with
\beq \label{phi0}
\phi_0 = \left({m_\phi M^{n-3}_{\rm P}\over
\lambda_n\sqrt{2n-2}}\right)^{1/(n-2)}\,.
\eeq
This results in Hubble expansion rate during inflation which is given by
\beq \label{hubble}
H_{\rm inf} = {(n-2) \over \sqrt{6 n (n-1)}} {m_{\phi} \phi_0 \over M_{\rm P}}.
\eeq
When $\phi$ is very close to $\phi_0$, the first derivative is
extremely small. The field is effectively in a de Sitter background,
and we are in self-reproduction (or {\it eternal inflation}) regime
where the two point correlation function for the flat direction
fluctuation grows with time. But eventually classical friction wins
and slow roll begins at $\phi \approx \phi_{\rm
self}$~\cite{AEGM,AEGJM}
\beq \label{self}
(\phi_0-\phi_{\rm self}) \simeq \Big({m_\phi \phi_0^2 \over M_{\rm
P}^3}\Big)^{1/2} \phi_0.
\eeq
The regime of {\it eternal inflation} plays an important role in
addressing the initial condition problem, see section 10.

The observationally relevant perturbations are generated when $\phi
\approx \phi_{\rm COBE}$. The number of e-foldings between $\phi_{\rm
COBE}$ and $\phi_{\rm end}$, denoted by ${\cal N}_{\rm COBE}$ follows
from Eq.~(\ref{efold})
\beq \label{cobe}
{\cal N}_{\rm COBE} \simeq {\phi^3_0 \over 2n(n-1)M^2_{\rm P}(\phi_0 -
\phi_{\rm COBE})}.
\eeq
The amplitude of perturbations thus produced is given by~\cite{AEGJM}
\beq \label{ampl}
\delta_{H} \equiv \frac{1}{5\pi}\frac{H^2_{\rm inf}}{\dot\phi} \simeq
\frac{1}{5\pi} \sqrt{\frac{2}{3}n(n-1)}(n-2) ~ \Big({m_\phi M_{\rm P} \over
\phi_0^2}\Big) ~ {\cal N}_{\rm COBE}^2,
\eeq
and the spectral tilt of the power spectrum and its running are found
to be~\cite{AEGM,AEGJM}
\begin{eqnarray}
\label{tilt}
&&n_s = 1 + 2\eta - 6\epsilon \ \simeq \ 1 -
{4\over {\cal N}_{\rm COBE}} \,, \\ \label{running}
&&{d\,n_s\over d\ln k} = - {4\over {\cal N}_{\rm COBE}^2}. \,
\end{eqnarray}
%

\subsection{Inflaton Properties and predictions}

\subsubsection{Inflaton candidates}

As discussed in~\cite{AEGM,AEGJM}, among nearly 300 flat directions
there are two that can lead to a successful inflation along the lines
discussed above.

One is $udd$ which, up to an overall phase factor, is parameterized by
\beq
\label{example}
u^{\alpha}_i=\frac1{\sqrt{3}}\phi\,,~
d^{\beta}_j=\frac1{\sqrt{3}}\phi\,,~
d^{\gamma}_k=\frac{1}{\sqrt{3}}\phi\,.
\eeq
Here $1 \leq \alpha,\beta,\gamma \leq 3$ are color indices, and $1
\leq i,j,k \leq 3$ denote the quark families. The flatness constraints
require that $\alpha \neq \beta \neq \gamma$ and $j \neq k$.

The other direction is $LLe$~\footnote{When the flat direction
develops a VEV during inflation, it spontaneously breaks $SU(2)\times
U(1)_{y}$, which gives masses to the corresponding gauge bosons.  It
is possible to obtain a seed perturbations for the primordial magnetic
field in this case, see~\cite{EJM}.}, parameterized by (again up to
an overall phase factor)
\beq
L^a_i=\frac1{\sqrt{3}}\left(\begin{array}{l}0\\ \phi\end{array}\right)\,,~
L^b_j=\frac1{\sqrt{3}}\left(\begin{array}{l}\phi\\ 0\end{array}\right)\,,~
e_k=\frac{1}{\sqrt{3}}\phi\,,
\eeq
where $1 \leq a,b \leq 2$ are the weak isospin indices and $1 \leq
i,j,k \leq 3$ denote the lepton families. The flatness constraints
require that $a \neq b$ and $i \neq j \neq k$.  Both these flat
directions are lifted by $n=6$ non-renormalizable
operators~\cite{gherghetta96},
\begin{eqnarray}
W_6\supset\frac{1}{M_{\rm P}^3}(LLe)(LLe)\,,\hspace{1cm}
W_6\supset \frac{1}{M_{\rm P}^3}(udd)(udd)\,.
\end{eqnarray}
The reason for choosing either of these two flat
directions\footnote{Since $LLe$ are $udd$q are independently $D$- and
$F$-flat, inflation could take place along any of them but also, at
least in principle, simultaneously. The dynamics of multiple flat
directions are however quite involved~\cite{Jokinen}.} is twofold: (i)
a non-trivial $A$-term arises, at the lowest order, only at $n=6$; and
(ii) we wish to obtain the correct COBE normalization of the CMB
spectrum.

Those MSSM flat directions which are lifted by operators with
dimension $n=7,9$ are such that the superpotential term contains at
least two monomials, i.e. is of the type
\begin{eqnarray}\label{doesnotcontri}
W \sim \frac{1}{M_{\rm P}^{n-3}}\Psi\Phi^{n-1}\,.
\end{eqnarray}
If $\phi$ represents the flat direction, then its VEV induces a large
effective mass term for $\psi$, through Yukawa couplings, so that
$\langle \psi \rangle =0$. Hence Eq. (\ref{doesnotcontri}) does not
contribute to the $A$-term.

More importantly, as we will see, all other flat directions except
those lifted by $n=6$ fail to yield the right amplitude for the
density perturbations. Indeed, as can be seen in Eq.~(\ref{phi0}), the
value of $\phi_0$, and hence also the energy density, depend on $n$.


\subsubsection{Inflaton Predictions}

According to the arguments presented above, successful MSSM flat direction
inflation has the following model parameters:
\beq
m_{\phi}\sim 1-10~{\rm TeV}\,,~~n=6\,,~~A=\sqrt{40}m_{\phi}\,,
~~\lambda\sim {\cal O}(1)\,.
\label{VALVS}
\eeq
Here we assume that $\lambda$ (we drop the subscript "6") is of order
one, which is the most natural assumption when $M=M_{\rm P}$.

The Hubble expansion rate during inflation and the VEV of the saddle
point are~\footnote{We note that $H_{\rm inf}$ and $\phi_0$ depend
very mildly on $\lambda$ as they are both $\propto \lambda^{-1/4}$.}
\beq \label{values}
H_{\rm inf}\sim 1-10~{\rm GeV}\,,~~~\phi_0 \sim (1-3) \times
10^{14}~{\rm GeV}\,.
\eeq
Note that both the scales are sub-Planckian. The total energy density
stored in the inflaton potential is $V_0 \sim 10^{36}-10^{38}~{\rm
GeV}^4$. The fact that $\phi_0$ is sub-Planckian guarantees that the
inflationary potential is free from the uncertainties about physics at
super-Planckian VEVs. The total number of e-foldings during the slow
roll evolution is large enough to dilute any dangerous relic
away~\cite{AEGJM},
\beq \label{totalefold}
{\cal N}_{\rm tot} \sim 10^3  \,,
\eeq
At such low scales as in MSSM inflation the number of e-foldings,
${\cal N}_{\rm COBE}$, required for the observationally relevant
perturbations, is much less than $60$~\cite{LEACH,BURGESS-multi}.  If
the inflaton decays immediately after the end of inflation, we obtain
${\cal N}_{\rm COBE} \sim 50$. Despite the low scale, the flat
direction can generate adequate density perturbations as required to
explain the COBE normalization. This is due to the extreme flatness of
the potential (recall that $V'=0$), which causes the velocity of the
rolling flat direction to be extremely small. From Eq.~(\ref{ampl}) we
find an amplitude of
\beq
\label{amp}
\delta_{H} \simeq 1.91 \times 10^{-5}\,.
\eeq

There is a constraint on the mass of the flat direction from the
amplitude of the CMB anisotropy:
\begin{equation}
\label{mbound} m_{\phi} \simeq (100 ~ {\rm GeV}) \times \lambda^{-1}
\, \left( \frac{{\cal N}_{\rm COBE}}{50} \right)^{-4}\,.
\end{equation}
We get a lower limit on the mass parameter when $\lambda\leq 1$.
For smaller values of $\lambda\ll 1$, the mass of the flat
direction must be larger.  Note that the above bound on the inflaton
mass arises at high scales, i.e. $\phi=\phi_0$. However, through
renormalization group flow, it is connected to the low scale mass, as
will be discussed in Sect. 4.

The spectral tilt of the power spectrum is not negligible because,
although the first slow roll parameter is $\epsilon\sim1/{\cal N}_{\rm
COBE}^4\ll 1$, the other slow roll parameter is given by $\eta =
-2/{\cal N}_{\rm COBE}$ and thus, see
Eq.~(\ref{tilt})\footnote{Obtaining $n_s > 0.92$ (or $n_s < 0.92$,
which is however outside the $2 \sigma$ allowed region) requires
deviation from the saddle point condition in Eq.~(\ref{cond}), see the
next subsection. For a more detailed discussion on the spectral tilt,
see also Refs.~\cite{LYTH1},\cite{AM}.}
\begin{eqnarray}
\label{spect}
&&n_s
\sim 0.92\,,\\
&&{d\,n_s\over d\ln k}
\sim - 0.002\,,
\end{eqnarray}
where we have taken ${\cal N}_{\rm COBE} \sim 50$ (which is the
maximum value allowed for the scale of inflation in our model). In the
absence of tensor modes, this agrees with the current WMAP 3-years'
data within $2\sigma$~\cite{WMAP3}. Note that MSSM inflation does not
produce any large stochastic gravitational wave background during
inflation. Gravity waves depend on the Hubble expansion rate, and in
our case the energy density stored in MSSM inflation is very small.


\subsection{Departure from the saddle point}

Inflation can still happen for small deviations from the saddle point
condition Eq.~(\ref{cond}). To quantify this, we define a parameter
$\alpha^2$ such that~\cite{AEGJM,AM}:
\beq \label{dev}
{A^2 \over 8 (n-1) m^2_{\phi}} \equiv 1 + \Big({n-2 \over 2}\Big)^2
\alpha^2\,.
\eeq
For $\alpha^2 \neq 0$, the saddle point becomes a point of inflection where
$V^{\prime \prime}(\phi_0) = 0$, and
\beq \label{1st}
V^{\prime}(\phi_0) = \Big({n-2 \over 2}\Big)^2 \alpha^2 m^2_{\phi}
\phi_0.
\eeq
If $\alpha^2 < 0$, the potential has a local minimum and a maximum.
In this case the flat direction is trapped in the local minimum. It
will eventually tunnel past the maximum and a period of slow roll
inflation will follow~\cite{AEGJM}. If $\alpha^2 > 0$, the potential
has no maximum or local minimum, and then slow roll inflation occurs
around $\phi_0$.

\begin{figure}[t]
\includegraphics[width=10.5cm]{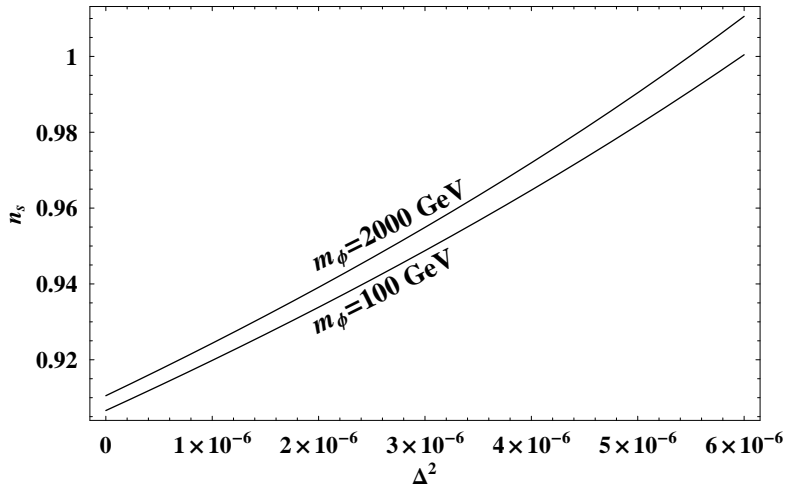}
\caption{$n_s$ is plotted as a function of $\Delta^2$ for different
values of $m_{\phi}$. $\Delta$ is defined in the text.  We choose
$\lambda$ =1.} \label{nsdel0}
\end{figure}

\begin{figure}[t]
\vspace{2cm}
\includegraphics[width=10.5cm]{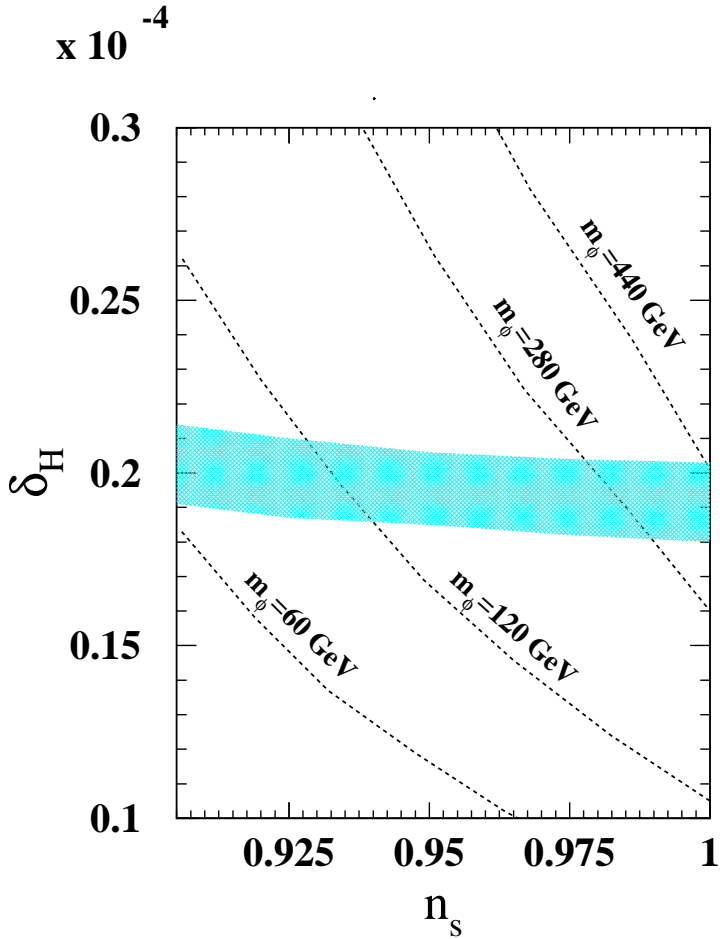}
\caption{$\delta_H$ is plotted as a function of $\Delta^2$ for
different values of $m_{\phi}$. We used $\lambda$ =1. The blue band
denotes the experimentally allowed values of $\delta_H$.}
\label{nsdel}
\end{figure}

For $\alpha^2 \neq 0$ the expressions for $n_s$ and $\delta_H$ are
modified as~\cite{LYTH1} (see also~\cite{AM})
\beq \label{ampl2} \delta_H = {1 \over 5 \pi} \sqrt{{2 \over 3}
n(n-1)} (n-2) {m_{\phi} M_{\rm P} \over \phi^2_0}{1 \over \Delta^2}
~ {\rm sin}^2 [{\cal N}_{\rm COBE}\sqrt{\Delta^2}]\,, \eeq
and
\beq \label{tilt2} n_s = 1 - 4 \sqrt{\Delta^2} ~ {\rm cot} [{\cal
N}_{\rm COBE}\sqrt{\Delta^2}],, \eeq
where
\beq \label{Delta} \Delta^2 \equiv n^2 (n-1)^2 \alpha^2 {\cal
N}^2_{\rm COBE} \Big({M_{\rm P} \over \phi_0}\Big)^4\,. \eeq
Note that for for $\alpha^2 = 0$, Eqs.~(\ref{ampl2},\ref{tilt2}) are
reduced to~(\ref{ampl},\ref{tilt}) respectively. For $\alpha^2 < 0$,
the spectral index will be smaller than that in Eq.~(\ref{tilt}),
thus outside the $2 \sigma$ region from observations. The more
interesting case, as pointed out in~\cite{AM}, happens for
$\alpha^2 > 0$.  We can in this case get all values within the
allowed range $0.92 \leq n_s \leq 1$~\cite{KKMR} for
\beq \label{Delta2} 0 \leq \Delta^2 \leq {\pi^2 \over 4 {\cal
N}^2_{\rm COBE}}\,. \eeq

The inflaton mass, $m_{\phi}$, is constrained by the experimental
data on the spectral index $n_s$~\cite{WMAP3,KKMR} and
$\delta_H$~\cite{liddle}.

We first find the solutions of $m_{\phi}$ by solving
Eqs.~(\ref{ampl2},\ref{tilt2}). $n_s$ depends mainly on $\Delta^2$ and
is mostly independent of $m_{\phi}$ and $\lambda$ (the coupling in
Eq.~(1)). The parameter $\Delta^2$ is defined in
Eq.~(\ref{Delta2}). We therefore solve $\Delta^2$ from
Eq.~(\ref{tilt2}) and apply this solution to determine the bounds on
$m_{\phi}$ from the Eq.~(\ref{ampl2}). In figure~\ref{nsdel0}, we show
$n_s$ as a function of $\Delta^2$. The range for $\Delta^2$ is
determined from Eq.~(\ref{Delta2}).\\

In figure~\ref{nsdel}, we show $\delta_H$ as a function of $n_s$ for
different values of $m_{\phi}$. The blue band shows the experimentally
allowed region. We find that smaller values of $m_{\phi}$ are
preferred for smaller values of $n_s$. We also find that the allowed
range of $m_{\phi}$ is $75-440$~GeV for the experimental ranges of
$n_s$ and $\delta_H$.  We assume $\lambda\sim 1$ for these two
figures. If $\lambda$ is less than ${\cal O}(1)$, e.g., $\lambda \sim
0.1$ or so (which can occur in $SO(10)$ model), it will lead to an
increase in $m_{\phi}$. We will need to study these allowed ranges of
the inflaton mass in the mSUGRA model. Since the inflaton mass is
related to the parameters of the mSUGRA model, the main question is
whether the allowed range of the inflaton mass is consistent with the
experimentally allowed mSUGRA model or not. See section 7.


\section{Radiative and supergravity corrections}

Since the MSSM inflaton candidates are represented by {\it gauge
invariant} combinations which are not singlets. The inflaton
parameters receive corrections from gauge interactions which, unlike
in models with a gauge singlet inflaton, can be computed in a
straightforward way.  Quantum corrections result in a logarithmic
running of the soft supersymmetry breaking parameters $m_\phi$ and
$A$. In this section we will discuss running of the potential with
VEV-dependent values of $m_\phi(\phi)$ and $A(\phi)$ in
Eq.~(\ref{cond}). Our conclusion is that the running of the gauge
couplings do not spoil the existence of a saddle point. However the
VEV of the saddle point is now displaced; by how much will depend
precisely on the inflaton candidate. In order to discuss the
situation, we derive a general expression for the one-loop effective
potential for the flat directions, and then focus on the $LLe$
direction, for which the system of Renormalization Group (RG)
equations can be solved analytically~\footnote{$udd$ case follows
similar discussion but requires numerics to solve the equations.}.

\subsection{One-loop effective potential}

The first thing to check is whether the radiative corrections remove
the saddle point altogether. The object of interest is the effective
potential at the phase minimum $n\theta_{\min} = \pi$, for which we
obtain~\cite{AEGJM}
\begin{eqnarray}
V_{eff}(\phi, \theta_{min}) &=& \frac{1}{2} m_0^2 \phi^2 \left[
  1+ K_1 \log \left(\frac{\phi^2}{\mu_0^2}\right) \right)] -
  \frac{\lambda_{n,0} A_0}{nM^{n-3}} \phi^n \left[ 1+ K_2
  \log\left(\frac{\phi^2}{\mu_0^2}\right) \right] \nonumber \\ & & +
  \frac{\lambda_{n,0}^2}{M^{2(n-3)}} \phi^{2(n-1)} \left[ 1 + K_3 \log
  \left(\frac{\phi^2}{\mu_0^2}\right) \right]\,.
\end{eqnarray}
%
%
%
%
%
where $m_0$, $A_0$, and $\lambda_{n,0}$ are the values of $m_{\phi}$,
$A$ and $\lambda_n$ given at a scale $\mu_0$. Here $A_0$ is chosen to
be real and positive (this can always be done by re-parameterizing the
phase of the complex scalar field $\phi$), and $|K_i|<1$ are
coefficients determined by the one-loop renormalization group
equations.

Our aim is to find a saddle point of this effective potential, so we
calculate the 1st and 2nd derivatives of the potential and set them to
zero. This is a straightforward although somewhat cumbersome exercise
that results in the expression~\cite{AEGJM}
\begin{equation}
\label{phi0-append}
\phi_0^{n-2} = \frac{M^{n-3}}{4\lambda_n(n-1+K_3)} \left[A
\left(
    1+ \frac{2}{n} K_2 \right) \pm \sqrt{A^2 \left( 1+ \frac{2}{n} K_2
    \right)^2 - 8m_{\phi}^2 (1+K_1) (n-1+K_3) } \right]\,,
\end{equation}
where $m_{\phi}$, $A$, and $\lambda_{n}$ are values of the parameters
at the scale $\phi_0$.  Inserting this into $V_{,\phi\phi}=0$, we can
then find the condition to have a saddle point at $\phi_0$:
\begin{eqnarray}
\label{A-append}
&&A^2 = 2 m_{\phi}^2 (n-1+K_3) F_1 F_2 F_3 \, \nonumber \\
&&F_1 = \left[ \frac{1+K_1}{n-1+K_3} \Big( (n-1)(2n-3) +
    (4n-5)K_3 \Big) - 1 - 3K_1 \right]^2 , \, \nonumber \\
&&F_2 = \left[ (1+K_1) \left(n-1+2\frac{2n-1}{n} K_2 \right) - (1+3K_1)
\left( 1+
    \frac{2}{n} K_2 \right) \right]^{-1}, \,  \nonumber \\
&&F_3 =
\left[
\frac{1+\frac{2}{n}K_2}{n-1+K_3} \Big( (n-1)(2n-3) + (4n-5)K_3 \Big)
- \left( n-1+ 2 \frac{2n-1}{n} K_2 \right) \right]^{-1}\,. \nonumber \\
&& \,
\end{eqnarray}
In the limit when $|K_i|\ll 1$, this mercifully simplifies to~\cite{AEGJM}
\begin{eqnarray} \label{A-append2}
A^2 = 8(n-1) m_{\phi}^2(\phi_0)
\left( 1+ K_1 - \frac{4}{n} K_2 + \frac{1}{n-1} K_3 \right)\,,\\
\phi_0^{n-2} = \frac{M^{n-3} m_{\phi}(\phi_0)}{\lambda_n \sqrt{2(n-1)}}
\left( 1+\frac{1}{2} K_1 - \frac{1}{2(n-1)} K_3 \right)\,.
\end{eqnarray}
Note that Eqs.~(\ref{A-append},\ref{A-append2}) give the necessary
relations between the values of $m_{\phi}$ and $A$ as calculated at
the {\it saddle point}.  The coefficients $K_i$ need to be solved from
the renormalization group equations at the scale given by the saddle
point $\mu=\phi_0$. Since $K_i$ are already one loop corrections,
taking the tree-level saddle point value as the renormalization scale
is sufficient.

The conclusion is robust, although the soft terms and the value of the
saddle point are all affected by radiative corrections, they do not
remove the saddle point nor shift it to unreasonable values. The
existence of a saddle point is thus insensitive to radiative
corrections.


\subsection{RG equations for the $LLe$ direction}

The form of the relevant RG equations depend on the flat direction. RG
equations for ${LLe}$ are simpler since only the $SU(2)_{W} \times
U(1)_Y$ gauge interactions are involved and the lepton Yukawa
couplings are negligible. The case of $udd$ requires numerics if
$u$ is chosen from the third family. For other choices, however, it
closely resembles ${LLe}$. For ${LLe}$ the one-loop RG
equations governing the running of $m^2_{\phi}$, $A$, and $\lambda$
with the scale $\mu$ are given by~\cite{Nilles}
\begin{eqnarray} \label{RGE}
\mu {d m^2_{\phi} \over d \mu} &=& - {1 \over 6 \pi^2} \left({3 \over 2}
{\tilde m_2}^2 g^2_2 + {3 \over 2} {\tilde m_1}^2 g^2_1 \right) \, , \nonumber
\\
\mu {d A \over d \mu} &=& - {1 \over 2 \pi^2} \left({3 \over 2} {\tilde m_2}
g^2_2 + {3 \over 2} {\tilde m_1} g^2_1 \right) \, , \nonumber \\
\mu {d \lambda \over d \mu} &=& - {1 \over 4 \pi^2} \lambda \left({3 \over 2}
g^2_2 + {3 \over 2} g^2_1 \right) \, .
\end{eqnarray}
Here ${\tilde m_1}$, ${\tilde m_2}$ denote the mass of the $U(1)_Y$
and $SU(2)_W$ gauginos respectively and $g_1,~g_2$ are the associated
gauge couplings. It is a straightforward exercise to obtain the
equations that govern the running of $\lambda$ and $A$ associated with
the $\left(LLe\right)^2$ superpotential term (which lifts the $LLe$
flat direction). Note that ${L}$ has the same quantum numbers as
${H}_d$, and hence in this respect ${LLe}$ combination behaves just
like ${H_{d}Le}$. One can then use the familiar RG equations that
govern the Yukawa coupling and $A$-term associated with the ${H}_d {L}
{e}$ superpotential term~\cite{Nilles}. However, as explained in
~\cite{Yamada}, the coefficients of the terms on the right-hand side
are proportional to the number of superfields contained in a
superpotential term.~\footnote{We would like to thank Manuel Drees for
explaining this point to us.} Hence the second and third equations
in~(\ref{RGE}) are simply obtained from those for the ${H}_d {Le}$
term after multiplying by a factor of $2$. The first equation
in~(\ref{RGE}) is also easily found by taking the electroweak charges
of $L_i$, $L_j$ and $e$ superfields into account while taking into
account that $m^2_{\phi} = (m^2_{L_i} + m^2_{L_j}+ m^2_{\bf e})/3$.

The running of gauge couplings and gaugino masses obey the usual
equations~\cite{Nilles}:
\begin{eqnarray}
\mu {d g_1 \over d \mu} &=& {11 \over 16 \pi^2} g^3_1 \, , \nonumber \\
\mu {d g_2 \over d \mu} &=& {1 \over 16 \pi^2} g^3_2 \, , \nonumber \\
{d \over d \mu}\left({{\tilde m_1} \over g_1^2}\right) &=&
{d \over d \mu}\left({{\tilde m_2} \over g_2^2}\right) = 0 \, .
\end{eqnarray}
The solutions of the RG equations are~\cite{AEGJM}
\begin{eqnarray}
g_i &=& \frac{g_i(\mu_0)}{\sqrt{1-b_i g_i(\mu_0)^2 \ln
    \frac{\mu}{\mu_0} }}, \\
{\tilde m_i} &=& {\tilde m_i}(\mu_0) \left( \frac{g_i}{g_i(\mu_0)}
\right)^2, \\
\label{mphi}
m_{\phi}^2 &=& m_{\phi}^2(\mu_0) + {\tilde m_2}^2(\mu_0) - {\tilde m_2}^2 +
\frac{1}{11} \left( {\tilde m_1}^2(\mu_0) - {\tilde m_1}^2 \right), \\
A &=& A(\mu_0) + 6 \left( {\tilde m_2}(\mu_0) - {\tilde m_2} \right) +
\frac{6}{11} \left( {\tilde m_1}(\mu_0) - {\tilde m_1} \right), \\
\lambda &=& \lambda(\mu_0) \left( \frac{g_2(\mu_0)}{g_2} \right)^6 \left(
  \frac{g_1(\mu_0)}{g_1} \right)^{\frac{6}{11}}~,
\end{eqnarray}
where $i=1,2$, $b_1=11/8\pi^2$ and $b_2=1/8\pi^2$.  Ignoring the
running of the gaugino masses and gauge couplings, we find
that~\cite{AEGJM}
\begin{eqnarray}
K_1 &\approx& - {1 \over 4 \pi^2} \left[\left({{\tilde m_2}
\over m_{\phi_0}}\right)^2 g^2_2 + \left({{\tilde m_1}
\over m_{\phi_0}}\right)^2 g^2_1 \right] \, , \nonumber \\
K_2 &\approx& - {3 \over 4 \pi^2} \left[\left({{\tilde m_2}
\over A_0}\right) g^2_2 +
\left({{\tilde m_1} \over A_0}\right) g^2_1 \right] \, , \nonumber \\
K_3 &\approx& - {3 \over 8 \pi^2} \lambda_0 \left[g^2_2 + g^2_1 \right] \, ,
\end{eqnarray}
where the subscript $0$ denotes the values of parameters at the high scale
$\mu_0$.

For universal boundary conditions, as in minimal grand unified
supergravity, the high scale is the GUT scale $\mu_X \approx 3 \times
10^{16}$~GeV, ${\tilde m_1}(\mu_X) = {\tilde m_2}(\mu_X) = {\tilde m}$
and $g_1 = \sqrt{\pi/10} \approx 0.56$, $g_2 = \sqrt{\pi/6} \approx
0.72$. Then we just use RG equations to run the coupling constants and
masses to the scale of the saddle point $\mu_0 = \phi_0 \approx 2.6
\times 10^{14}$~GeV for $M_P = 2.4 \times 10^{18}$~GeV, $m_{\phi_0}=
1$~TeV, $\lambda_0=1$. With these values we obtain~\cite{AEGJM}
\begin{eqnarray}
K_1 &\approx & -0.017 \xi^2, \\
K_2 &\approx & -0.0085 \xi, \\
K_3 &\approx & -0.029\,.
\end{eqnarray}
where $\xi = {\tilde m}/ m_{\phi}$ is calculated at the GUT scale.

Typically the running based on gaugino loops alone results in negative
values of $K_i$~\cite{EJM1}. Positive values can be obtained when one
includes the Yukawa couplings, practically the top Yukawa, but the
order of magnitude remains the same.


\begin{figure}
\vspace*{-0.0cm}
\begin{center}
\epsfig{figure=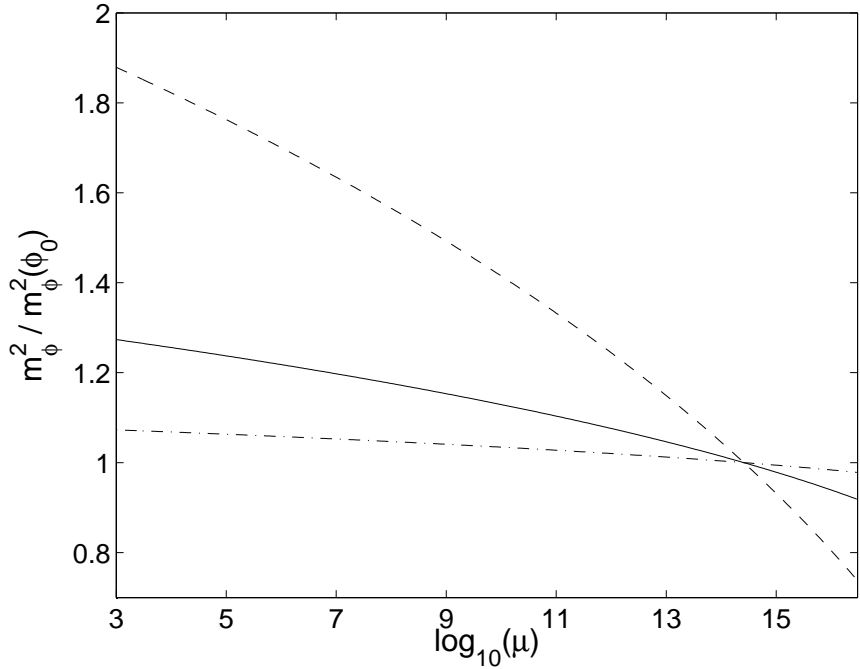,width=.84\textwidth,clip=}
\vspace*{-0.0cm}
\end{center}
\caption{The running of $m^2_{\phi}$ for the $LLe$ inflaton when the
saddle point is at $\phi_0 = 2.6 \times 10^{14}$GeV (corresponding to
$n=6$, $m_{\phi}=1$ TeV and $\lambda = 1$). The three curves
correspond to different values of the ratio of gaugino mass to flat
direction mass at the GUT scale: $\xi = 2$ (dashed), $\xi = 1$ (solid)
and $\xi = 0.5$ (dash-dot).  }\label{LLE-RUN0}
\end{figure}


Thus radiative corrections modify $\alpha$ and we need to fine tune the
potential to a few (but not all) orders in perturbation theory.


\subsection{ $A_6$~vs.~$A_3$}

One final comment is in order before closing this Section. Unlike
$m_{\phi}$, there is no prospect of measuring the $A$ term, because it
is related to the non-renormalizable interactions which are suppressed
by $M_{\rm P}$. However, a knowledge of supersymmetry breaking sector
and its communication with the observable sector may help to link the
non-renormalizable $A$-term under consideration to the renormalizable
ones.

To elucidate this, let us consider the Polonyi model where a general
$A$-term at a tree level is given by
$$m_{3/2}[(a-3)W+\phi (dW/d\phi)],$$
with $a=3 - \sqrt{3}$~\cite{Nilles}. One then finds a relationship
between $A$-terms corresponding to $n=6$ and $n=3$ superpotential
terms, denoted by $A_6$ and $A_3$ respectively, at high scales:
\beq \label{polon}
A_6={3 - \sqrt{3} \over 6 - \sqrt{3}} A_3\,.
\eeq
One can then use relevant RG equations to relate $A_6$ which is
relevant for inflation, to $A_3$ at the weak scale, which can be
constrained and/or measured. In principle this can also be done in
general, provided that we have sufficient information about the
supersymmetry breaking sector and its communication with the MSSM
sector, see some related discussions in~\cite{Enqvist-new}.


\subsection{Supergravity corrections}

SUGRA corrections often destroy the slow roll predictions of
inflationary potentials; this is the notorious SUGRA-$\eta$
problem~\cite{ETA}. In general, the effective potential depends on the
K\"ahler potential $K$ as $ V\sim
\left(e^{K(\varphi^{\ast},\varphi)/M_{\rm P}^2} V(\phi)\right) $ so
that there is a generic SUGRA contribution to the flat direction
potential of the type
\begin{equation}
\label{mflat}
V(\phi)=H^2M_{\rm P}^2 f\left(\frac{\phi}{M_{\rm P}}\right)\,,
\end{equation}
where $f$ is some function (typically a polynomial).  Such a
contribution usually gives rise to a Hubble induced correction to the
mass of the flat direction with an unknown coefficient, which depends
on the nature of the K\"ahler potential~\footnote{If the K\"ahler
potential has a shift symmetry, then at tree level there is no Hubble
induced correction. However, at one-loop level relatively small Hubble
induced corrections can be induced~\cite{GMO,ADM1}.}.

Let us compare the non-gravitational contribution, Eq.~(\ref{scpot}),
to that of Hubble induced contribution, Eq.~(\ref{mflat}). Writing
$f\sim \left( \phi/M_{\rm P}\right)^p$ where $p\ge 1$ is some power,
we see that non-gravitational part dominates whenever
\beq
H_{\rm inf}^2M_{\rm P}^2\left(\frac{\phi}{M_{\rm P}}\right)^p \ll
m_{\phi}^2\phi_0^2\,,
\eeq
so that the SUGRA corrections are negligible as long as $\phi_0 \ll
M_{\rm P}$, as is the case here (note that $H_{\rm inf} M_{\rm P} \sim
m_{\phi} \phi_0$).  The absence of SUGRA corrections is a generic
property of this model. Note also that although non-trivial K\"ahler
potentials give rise to non-canonical kinetic terms of squarks and
sleptons, it is a trivial exercise to show that at sufficiently low
scales, $H_{\rm inf}\ll m_{\phi}$, and small VEVs, they can be rotated
to a canonical form without affecting the potential~\footnote{The same
reason, i.e. $H_{\rm inf}\ll m_{\phi}$ also precludes any large
Trans-Planckian correction. Any such correction would generically go
as $(H_{\rm inf}/M_{\ast})^2\ll 1$, where $M_{\ast}$ is the scale at which
one would expect Trans-Planckian effects to kick in.}.


\section{Reheating and Thermalization}

After the end of inflation, the flat direction starts rolling towards
its global minimum. At this stage the dominant term in the scalar
potential will be: $m_\phi \phi^2/2$. Since the frequency of
oscillations is $\omega \sim m_{\phi} \sim 10^3 H_{\rm inf}$, the flat
direction oscillates a large number of times within the first Hubble
time after the end of inflation. Hence the effect of expansion is
negligible.

We recall that the curvature of the potential along the angular
direction is much larger than $H^2_{\rm inf}$. Therefore, the flat
direction has settled at one of the minima along the angular direction
during inflation from which it cannot be displaced by quantum
fluctuations. This implies that no torque will be exerted, and hence
the flat direction motion will be one dimensional, i.e. along the
radial direction.

Flat direction oscillations excite those MSSM degrees of freedom which
are coupled to it.  The inflaton, either ${LLe}$ or ${udd}$ flat
direction, is a linear combination of slepton or squark
fields. Therefore inflaton has gauge couplings to the gauge/gaugino
fields and Yukawa couplings to the Higgs/Higgsino fields. As we will
see particles with a larger couplings are produced more copiously
during inflaton oscillations. Therefore we focus on the production of
gauge fields and gauginos. Keep in mind that the VEV of the MSSM flat
direction breaks the gauge symmetry spontaneously, for instance ${\bf
udd}$ breaks $SU(3)_C \times U(1)_Y$ while ${\bf LLe}$ breaks
$SU(2)_{W}\times U(1)_{Y}$, therefore, induces a supersymmetry
conserving mass $\sim g \langle \phi(t) \rangle$ to the gauge/gaugino
fields in a similar way as the Higgs mechanism, where $g$ is a gauge
coupling. When the flat direction goes to its minimum, $\langle
\phi(t)\rangle = 0$, the gauge symmetry is restored. In this respect
the origin is a point of enhanced symmetry~\cite{AVERDI2}.

There can be various phases of particle creation in this model, here
we briefly summarize them below.  Let us elucidate the physics, by
considering the case when ${LLe}$ flat direction is the inflaton
~\footnote{ Reheating happens quickly due to a flat direction motion
which is {\it strictly} one dimensional in our case. Our case is
really exceptional, usually, the flat direction motion is restricted
to a plane, which precludes preheating all together, for instance
see~\cite{Rouz-Camp,Postma,Longeviety}.}.

\begin{itemize}

\item{Tachyonic preheating:\\ Right after the end of inflation, when
we are close to the saddle point, the second derivative is negative.
One might suspect that this would trigger tachyonic instability in the
inflaton fluctuations which will then excite the inflaton couplings to
matter~\cite{NEWPREH,ESTER}.

However the situation is different in our case. As mentioned, only
inflaton fluctuations with a physical momentum $k \ls m_{\phi}$ will
have a tachyonic instability.  Moreover $V^{\prime \prime} < 0$ only
at field values which are $\sim \phi_0$.  Tachyonic effects are
therefore expected be negligible since, unlike the case
in~\cite{NEWPREH}, the homogeneous mode has a VEV which is
hierarchically larger than $m_{\phi}$ (we remind that $\phi_0 \geq
10^{14}$ GeV) and oscillates at a frequency $\omega \sim
m_{\phi}$. Further note fields which are coupled to the inflaton
acquire a very large mass $\sim h \phi_0$ from the homogeneous piece
which suppresses non-perturbative production of their quanta at large
inflaton VEVs. We conclude that tachyonic effects, although genuinely
present, do not lead to significant particle production in our case.
}

\item{Instant preheating:\\ An efficient bout of particle creation
occurs when the inflaton crosses the origin, which happens twice in
every oscillation. The reason is that fields which are coupled to the
inflaton are massless near the point of enhanced symmetry. Mainly
electroweak gauge fields and gauginos are then created as they have
the largest coupling to the flat direction. The production takes place
in a short interval, $\Delta t \sim \left(g m_{\phi} \phi_0
\right)^{-1/2}$, where $\phi_0\sim 10^{14}$~GeV is the initial
amplitude of the inflaton oscillation, during which quanta with a
physical momentum $k \ls \left(g m_{\phi} \phi_0 \right)^{1/2}$ are
produced. The number density of gauge/gaugino degrees of freedom is given
by~\cite{PREHEAT}, see also~\cite{Cormier}
\beq \label{chiden}
n_{g} \approx {\left(g m_{\phi} \phi_0
\right)^{3/2} \over 8 \pi^3}\,.
\eeq
As the inflaton VEV is rolling back to its maximum value $\phi_0$, the
mass of the produced quanta $g \langle \phi(t) \rangle$ increases. The
gauge and gaugino fields can (perturbatively) decay to the fields
which are not coupled to the inflaton, for instance to (s)quarks. Note
that (s)quarks are not coupled to the flat direction, hence they
remain massless throughout the oscillations. The total decay rate of
the gauge/gaugino fields is then given by $\Gamma = C \left(g^2/48\pi
\right) g\phi $, where $C \sim {\cal O}(10)$ is a numerical factor
counting for the multiplicity of final states.

The decay of the gauge/gauginos become efficient when~\cite{AEGJM}
\beq \label{ddecay}
\langle \phi \rangle \simeq \left({48 \pi m_{\phi} \phi_0 \over
C g^3}\right)^{1/2}\,.
\eeq
Here we have used $\langle \phi(t) \rangle \approx \phi_0 m_{\phi} t$,
which is valid when $m_{\phi} t \ll 1$, and $\Gamma \simeq t^{-1}$,
where $t$ represents the time that has elapsed from the moment that
the inflaton crossed the origin. Note that the decay is very quick
compared with the frequency of inflaton oscillations, i.e. $\Gamma \gg
m_{\phi}$. It produces relativistic (s)quarks with an
energy~\cite{AEGJM}:
\beq \label{energy}
E =\frac{1}{2}g\phi(t)
\simeq \left({48 \pi m_{\phi} \phi_0 \over C g}\right)^{1/2}\,.
\eeq
The ratio of energy density in relativistic particles thus produced
$\rho_{rel}$with respect to the total energy density $\rho_0$ follows
from Eqs.~(\ref{chiden},\ref{energy}):
\beq
\label{ratio}
{\rho_{rel} \over \rho_0} \sim 10^{-2} g\,,
\eeq
where we have used $C \sim {\cal O}(10)$.  This implies that a
fraction $\sim {\cal O}(10^{-2})$ of the inflaton energy density is
transferred into relativistic (s)quarks every time that the inflaton
passes through the origin. This is so-called instant preheating
mechanism~\cite{INSTANT}~\footnote{In a favorable condition the flat
direction VEV coupled very weakly to the flat direction inflaton could
also enhance the perturbative decay rate of the
inflaton~\cite{ABM}.}. It is quite an efficient mechanism in our model
as it can convert almost all of the energy density in the inflaton into
radiation within a Hubble time (note that $H^{-1}_{\rm inf} \sim 10^3
m^{-1}_{\phi}$).}

\end{itemize}

\subsection{Towards thermal equilibrium}

A full thermal equilibrium is reached when ${\it a)~kinetic }$ and
${\it b)~chemical~equilibrium }$ are established. The maximum
(hypothetical) temperature attained by the plasma would be given by:
\beq
\label{tmax}
T_{max} \sim V^{1/4} \sim \left(m_{\phi}\phi_0\right)^{1/2}
\geq 10^{9}~{\rm GeV}\,.
\eeq
This temperature may be too high and could lead to thermal
overproduction of gravitinos~\cite{Ellis,Buchmuller}. However the
dominant source of gravitino production in a thermal bath is
scattering which include an on-shell gluon or gluino leg. In the next
subsection we describe a natural solution to this problem and show that
 the final reheat temperature is actually well below
Eq.~(\ref{tmax}), i.e. $T_{R}\ll T_{max}$.

One comment is in order before closing this subsection. The gravitinos
can also be created non-perturbatively during inflaton oscillations,
both of the helicity $\pm 3/2$~\cite{MAROTO} and helicity $\pm 1/2$
states~\cite{REST}. In models of high scale inflation (i.e. $H_{\rm
inf} \gg m_{3/2}$) helicity $\pm 1/2$ states can be produced very
efficiently (and much more copiously than helicity $\pm 3/2$ states).
At the time of production these states mainly consist of the inflatino
(inflaton's superpartner).  However these fermions also decay in the
form of inflatino, which is coupled to matter with a strength which is equal
to that of the inflaton. Therefore, they inevitably decay at a similar
rate as that of inflaton, and hence pose no threat to primordial
nucleosynthesis~\cite{MAR}.

In the present case $m_{\phi} \sim m_{3/2} \gg H_{\rm inf}$. Therefore
low energy supersymmetry breaking is dominant during inflation, and
hence helicity $\pm 1/2$ states of the gravitino are not related to
the inflatino (which is a linear combination of leptons or quarks)at
any moment of time. As a result helicity $\pm 1/2$ and $\pm 3/2$
states are excited equally, and their abundances are suppressed due to
kinematical phase factor.  Moreover there will be no dangerous
gravitino production from perturbative decay of the inflaton
quanta~\cite{AVERDI1,AVERDI3,SURFACE}. The reason is that the
inflaton is not a {\it gauge singlet} and has gauge strength couplings
to other MSSM fields. This makes the $inflaton \rightarrow inflatino
~+~ gravitino$ ~decay mode totally irrelevant.


\subsection{Solution to the gravitino problem}

In order to suppress thermal gravitino production it is sufficient to
make gluon and gluino fields heavy enough such that they are not
kinematically accessible to the reheated plasma, even if other degrees
of freedom reach full equilibrium (for a detailed discussion on
thermalization in supersymmetric models and its implications,
see~\cite{AVERDI1,AVERDI2}). This suggests a natural solution to the
thermal gravitino problem in the case of our model. Consider another
flat direction with a non-zero VEV, denoted by $\varphi$, which
spontaneously breaks the $SU(3)_C$ group.  For example, if ${LLe}$ is
the inflaton, then ${udd}$ provides a unique candidate which can
simultaneously develop VEV~\footnote{To develop and maintain such a
large VEV, it is not necessary that ${udd}$ potential has a saddle
point as well. It can be trapped in a false minimum during inflation,
which will then be lifted by thermal corrections when the inflaton
decays (as discussed in the previous subsection)~\cite{AEJM}.}The
induced mass for gluon/gluino fields will be:
\beq \label{flatvev}
m_{G} \sim  g \langle \varphi(t) \rangle < g \phi_0\,.
\eeq
The inequality arises due to the fact that the VEV of $\varphi$ cannot
exceed that of the inflaton $\phi$ since its energy density should be
subdominant to the inflaton energy density.

If $g \varphi_0 \gg T_{max}$ the gluon/gluino fields will be too heavy
and not kinematically accessible to the reheated plasma. Here
$\varphi_0$ is the VEV of ${\bf udd}$ at the beginning of inflaton
oscillations.  In a radiation-dominated Universe the Hubble expansion
redshifts the flat direction VEV as $\langle \varphi \rangle \propto
H^{3/4}$, which is a faster rate than the change in the temperature $T
\propto H^{1/2}$. Once $g \langle \varphi \rangle \simeq T$,
gluon/gluino fields come into equilibrium with the thermal bath.  As
pointed out in Refs.~\cite{AVERDI1,AVERDI2,AEGJM}, if the initial VEV
of ${udd}$ is
\beq
\label{reqVEV}
\varphi_0 > 10^{10}~{\rm GeV}\,,
\eeq
then the temperature at which gluon/gluino become kinematically
accessible, i.e. $g \langle \varphi \rangle \simeq T$, is given by
~\cite{AVERDI2}~\footnote{Note that the conditions in
Eqs.~(\ref{flatvev},\ref{reqVEV}) can be simultaneously satisfied
easily.}:
\beq
T_{\rm R} \leq 10^{7}~{\rm GeV}\,.
\eeq
This is the final reheat temperature at which gluons and gluinos are
all in thermal equilibrium with the other degrees of freedom. The
standard calculation of thermal gravitino production via scatterings
can then be used for $T \leq T_{\rm R}$. Note however that $T_{\rm R}$
is sufficiently low to avoid thermal overproduction of gravitinos.

Finally, we also make a comment on the cosmological moduli problem.
The moduli are generically displaced from their true
minimum if their mass is less than the expansion rate during
inflation. The moduli obtain a mass $\sim {\cal O}({\rm TeV})$ from
supersymmetry breaking. They start oscillating with a large amplitude,
possibly as big as $M_{\rm P}$, when the Hubble parameter drops below
their mass. Since moduli are only gravitationally coupled to other
fields, their oscillations dominate the Universe while they decay very
late.  The resulting reheat temperature is below MeV, and is too low
to yield a successful primordial nucleosynthesis.

However, in our case $H_{\rm inf} \ll {\rm TeV}$ . This implies that
quantum fluctuations cannot displace the moduli from their true minima
during the inflationary epoch driven by MSSM flat
directions. Moreover, any oscillations of the moduli will be
exponentially damped during the inflationary epoch. Therefore our
model is free from the infamous moduli problem~\cite{AEGJM}.


\section{Cold dark matter and MSSM inflation}

Since $m_{\phi}$ is related to the scalar masses, sleptons ($LLe$
direction) and squarks ($udd$ direction), the bound on $m_{\phi}$
will be translated into the bounds on these scalar masses which are
expressed in terms of the model parameters~\cite{AEGJM}. 

The models of mSUGRA depend only on four parameters and one
sign. These are $m_0$ (the universal scalar soft breaking mass at the
GUT scale $M_{\rm G}$); $m_{1/2}$ (the universal gaugino soft breaking
mass at $M_{\rm G}$); $A_0$ (the universal trilinear soft breaking
mass at $M_{\rm G}$)~\footnote{The relationship between the two $A$
terms, the trilinear, $A_0$ and the non-renormalizable $A$ term in
Eq.(\ref{scpot}) can be related to each other, however, that depends
on the SUSY breaking sector. For a Polonyi model, they are given by:
$A=(3-\sqrt{3})/(6-\sqrt{3})A_0$~\cite{AEGJM}.}; $\tan\beta = \langle
H_2 \rangle \langle H_1 \rangle$ at the electroweak scale (where $H_2$
gives rise to $u$ quark masses and $H_1$ to $d$ quark and lepton
masses); and the sign of $\mu$, the Higgs mixing parameter in the
superpotential ($W_{\mu} = \mu H_1 H_2$).  Unification of gauge
couplings within supersymmetry suggests that $M_{\rm G} \simeq 2
\times 10^{16}$ GeV. The model parameters are already significantly
constrained by different experimental results.  Most important
constraints are:

\begin{itemize}

\item{The light Higgs mass bound of $M_{h^0} > 114.0$~GeV from LEP
\cite{LEP}.}

\item{The $b \rightarrow s \gamma$ branching ratio~\cite{bsgamma}:
$2.2\times10^{-4} < {\cal B}(B \rightarrow X_s \gamma) <
4.5\times10^{-4}$.}

\item{In mSUGRA the $\tilde\chi^0_1$ is the candidate for CDM.  The
$2\sigma$ bound from the WMAP \cite{WMAP3} gives a relic density bound
for CDM to be $0.095 < \Omega_{\rm CDM} h^2 < 0.129 $.}

\item{The bound on the lightest chargino mass of
$M_{\tilde\chi^{\pm}_1} > 104$~GeV from LEP~\cite{aleph}.}

\item{The possible $3.3~\sigma$ deviation (using $e^+e^-$ data to
calculate the leading order hadronic contribution)from the SM
expectation of the anomalous muon magnetic moment from the muon $g-2$
collaboration \cite{BNL}.}

\end{itemize}

\begin{figure}[t]
\vspace{1cm} \center
\includegraphics[width=10.0cm]{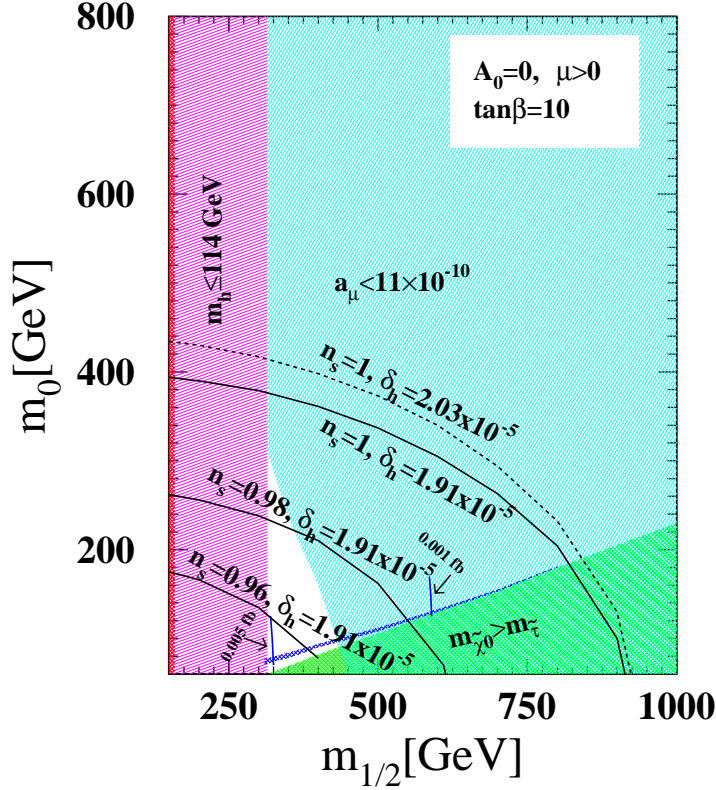} \caption{The
contours for different values of $n_s$ and $\delta_H$ are shown in
the $m_0-m_{1/2}$ plane for $\tan\beta=10$. We used $\lambda=1$ for
the contours. We show the dark matter allowed region {narrow blue
corridor}, (g-2)$_\mu$ region (light blue) for $a_{\mu}\leq
11\times10^{-8}$, Higgs mass $\leq 114$ GeV (pink region) and LEPII
bounds on SUSY masses (red). We also show the the dark matter
detection rate by vertical blue lines.} \label{10flat}
\end{figure}


\begin{figure}[t]
\vspace{1cm} \center \includegraphics[width=10.0cm]{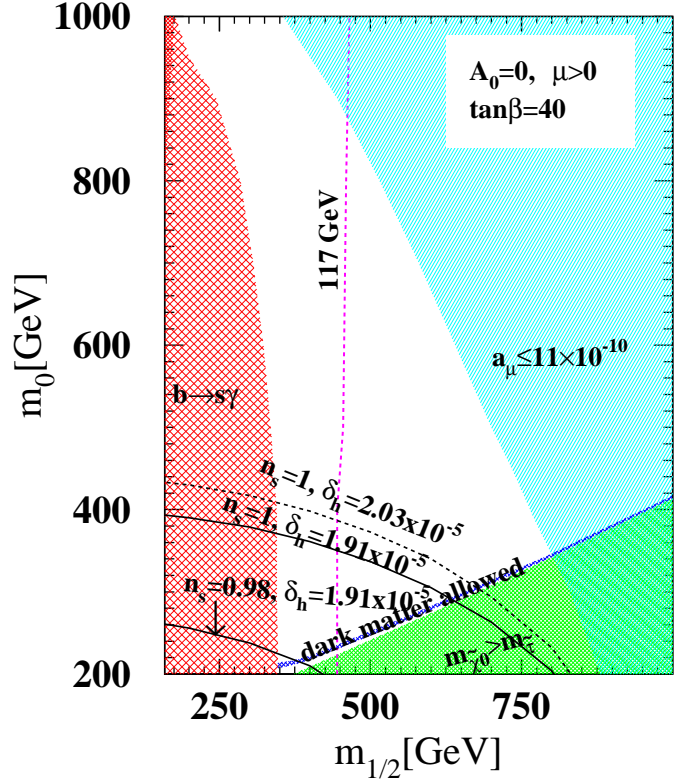}
\caption{The contours for different values of $n_s$ and $\delta_H$
are shown in the $m_0-m_{1/2}$ plane for $\tan\beta=40$. We used
$\lambda=1$ for the contours. We show the dark matter allowed region
{narrow blue corridor}, (g-2)$_\mu$ region (light blue) for
$a_{\mu}\leq 11\times10^{-8}$, $b\rightarrow s\gamma $ allowed
region (brick) and LEPII bounds on SUSY masses (red).}
\label{40flat}
\end{figure}

\begin{figure}[t]
\vspace{1cm} \center
\includegraphics[width=10.5cm]{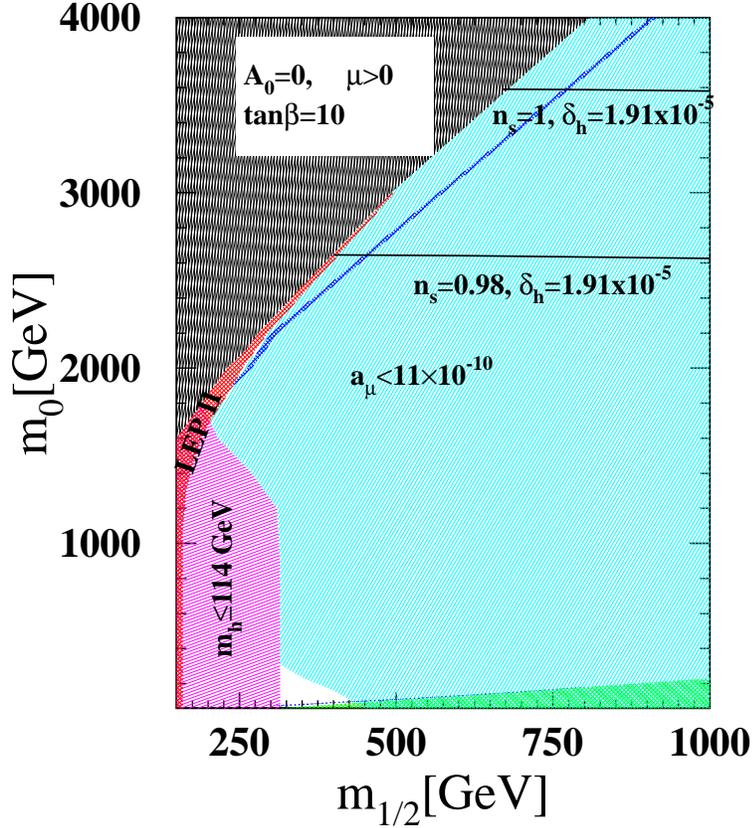}
\caption{The contours for different values of $n_s$ and $\delta_H$
are shown in the $m_0-m_{1/2}$ plane for $\tan\beta=10$. We used
$\lambda=0.1$ for the contours. We show the dark matter allowed
region {narrow blue corridor}, g-2 region (light blue) for
$a_{\mu}\leq 11\times10^{-8}$, Higgs mass $\leq 114$ GeV (pink
region) and LEPII bounds on SUSY masses (red). The black region is
not allowed by radiative electroweak symmetry breaking. We use
$m_t=172.7$~GeV for this graph.} \label{10flatfocus}
\end{figure}


\begin{figure}[t]
\vspace{1cm}
\includegraphics[width=10.5cm]{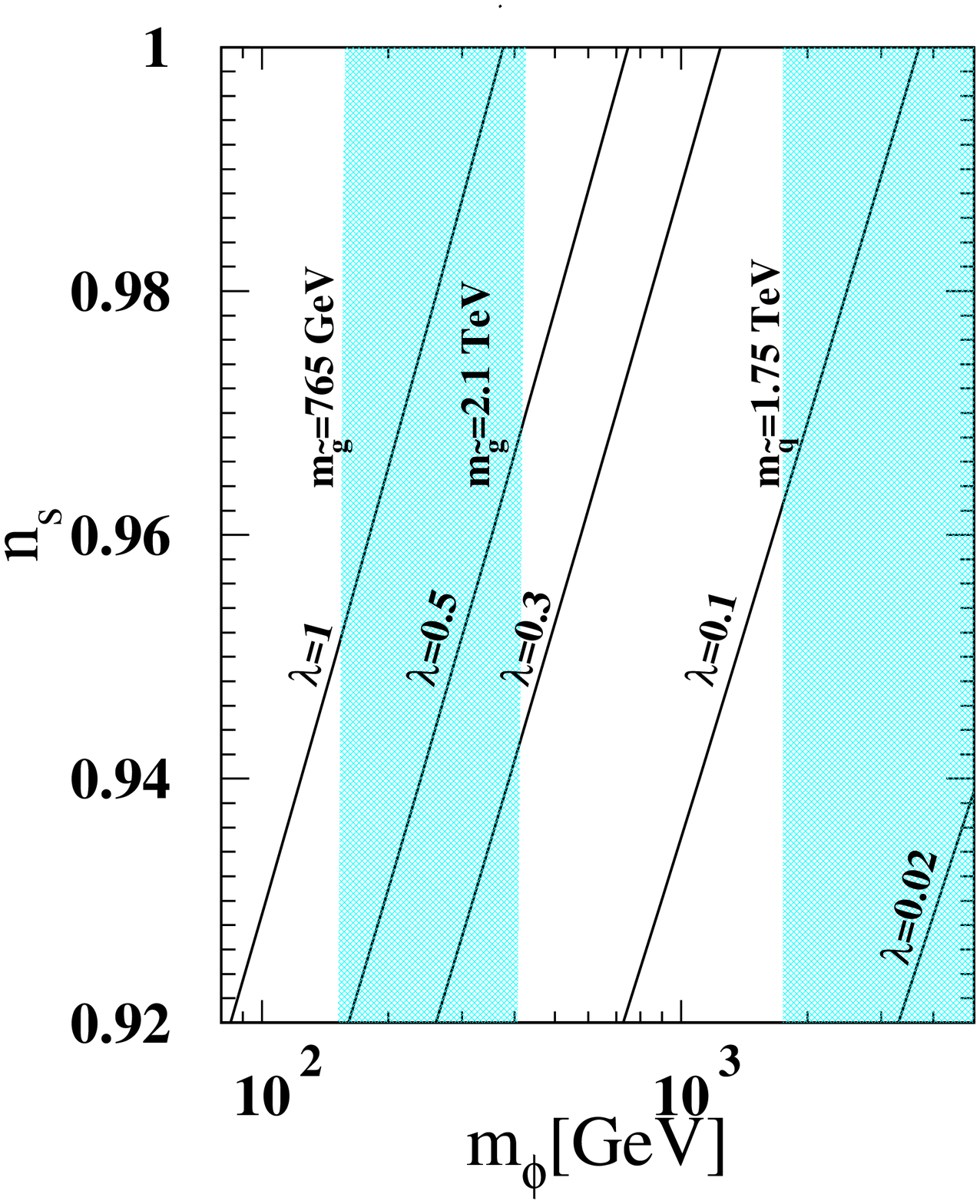}
\caption{Contours of $\lambda$  for $\delta_H=1.91\times 10^{-5}$ in
the $n_s$-$m_{\phi}$ plane. The blue band on the left is due to the
stau-neutralino coannihilation region for $\tan\beta=10$ and the
blue band on the right (which continues beyond the plotting range)
denotes the focus point region.} \label{lamcon}
\end{figure}


The allowed mSUGRA parameter space, at present, has mostly three
distinct regions: (i)~the stau-neutralino
($\tilde\tau_1~-~\tilde\chi^1_0$), coannihilation region where
$\tilde\chi^1_0$ is the lightest SUSY particle (LSP), (ii)~the
$\tilde\chi^1_0$ having a dominant Higgsino component (focus point)
and (iii)~the scalar Higgs ($A^0$, $H^0$) annihilation funnel
(2$M_{\tilde\chi^1_0}\simeq M_{A^0,H^0}$). These three regions have
been selected out by the CDM constraint. There stills exists a bulk
region where none of these above properties is observed, but this
region is now very small due to the existence of other experimental
bounds. After considering all these bounds we will show that there
exists an interesting overlap between the constraints from inflation
and the CDM abundance~\cite{ADM}.

We calculate $m_{\phi}$ at $\phi_0$ and $\phi_0$ is $~10^{14}$ GeV
which is two orders of magnitude below the GUT scale. From this
$m_{\phi}$, we determine $m_0$ and $m_{1/2}$ by solving the RGEs for
fixed values of $A_0$ and $\tan\beta$. The RGEs for $m_{\phi}$ are
\begin{eqnarray}
\mu{dm_{\phi}^2\over{d\mu}}&=&{-1\over{6\pi^2}}({3\over
2}{M_2^2}g_2^2+{9\over {10}}{M_1^2}g_1^2)\,,\quad\, ({\rm for\,
LLe})\, \nonumber \\
\mu{dm_{\phi}^2\over{d\mu}}&=&{-1\over{6\pi^2}}({4}{M_3^2}g_3^2+
{2\over {5}}{M_1^2}g_1^2)\,, \quad\, ({\rm for\, udd})\,.
\end{eqnarray}
$M_1,~M_{2}$ and $M_3$ are $U(1),~SU(2)$ and $SU(3)$ gaugino masses
respectively.

After we determine $m_0$ and $m_{1/2}$ from $m_{\phi}$, we can
determine the allowed values of $m_{\phi}$ from the experimental
bounds on the mSUGRA parameters space. In order to obtain the
constraint on the mSUGRA parameter space, we calculate the SUSY
particle masses by solving the RGEs at the weak scale using four
parameters of the mSUGRA model and then use these masses to
calculate Higgs mass, $BR[b\rightarrow s \gamma]$, dark matter
content etc.

We show that the mSUGRA parameter space in figures~\ref{10flat},
\ref{40flat} for $\tan\beta=10$ and $40$ with the $udd$ flat
direction using $\lambda=1$~\footnote{We have a similar figure for
the flat direction $LLe$ which we do not show in this paper. All the
figures are for $udd$ flat direction as an inflaton.}. In the
figures, we show contours correspond to $n_s=1$ for the maximum
value of $\delta_H=2.03\times 10^{-5}$ (at $2\sigma$ level) and
$n_s=1.0,~0.98,~0.96$ for $\delta_H=1.91\times 10^{-5}$. The
constraints on the parameter space arising from the inflation
appearing to be consistent with the constraints arising from the
dark matter content of the universe and other experimental results.
 We find that $\tan\beta$ needs to
be smaller to allow for smaller values of $n_s<1$. It is also
interesting to note that the allowed region of $m_{\phi}$, as
required by the inflation data for $\lambda=1$ lies in the
stau-neutralino coannihilation region which requires smaller values
of the SUSY particle masses. The SUSY particles in this parameter
space are, therefore, within the reach of the LHC very quickly. The
detection of the region at the LHC has been considered in
refs~\cite{dka}. From the figures, one can also find that as
$\tan\beta$ increases, the inflation data along with the dark
matter, rare decay and Higgs mass constraint  allow smaller ranges
of $m_{1/2}$. For example, the allowed ranges of  gluino masses are
765 GeV-2.1 TeV and 900 GeV-1.7 TeV for $\tan\beta=10$ and 40
respectively~\cite{ADM}.

So far we have chosen $\lambda=1$. Now if $\lambda$ is small e.g.,
$\lambda\ls 10^{-1}$, we find that the allowed values of $m_{\phi}$
to be large. In this case the dark matter allowed region requires
the lightest neutralino to have larger Higgsino component in the
mSUGRA model. As we will see shortly, this small value of $\lambda$
is accommodated in $SO(10)$ type model. In figure~\ref{10flatfocus},
we show $n_s=1,\, 0.98$ contours for $\delta_H=1.91\times 10^{-5}$
in the mSUGRA parameter space for $\tan\beta=10$. In this figure, we
find that $n_s$ can not smaller than 0.97, but if we lower $\lambda$
which will demand larger $m_{\phi}$ and therefore $n_s$ can be
lowered down to 0.92.

In figure~\ref{lamcon}, we show the contours of $\lambda$ for
different values of $m_{\phi}$ which are allowed by $n_s$ and
$\delta_H=1.91\times 10^{-3}$. The blue bands show the dark matter
allowed regions for $\tan\beta=10$. The band on the left is due to
the stau-neutralino coannihilation region allowed by other
constraints and the allowed values of $\lambda$ are 0.3-1.  The
first two generation squarks masses are 690 GeV and 1.9 TeV for  the
minimum and maximum values of $m_{\phi}$ allowed by the dark matter
and other constraints. The gluino masses for these are $765$~GeV and
$2.1$~TeV respectively.  The band is slightly curved due to the
shifting of $\phi_0$ as a function $\lambda$. (We solve for SUSY
parameters from the inflaton mass at $\phi_0$). The band on the
right which continues beyond the plotting range of the
figure~\ref{lamcon} is due to the Higgsino dominated dark matter. We
find that $\lambda$ is mostly $\leq 0.1$ in this region and
$m_{\phi}>1.9$ TeV. In this case the squark masses are much larger
than the gluino mass since $m_0$ is much larger than $m_{1/2}$.


\section{Grand unified Models and Inclusion of Right-Handed Neutrinos}

\subsection{Embedding MSSM inflation in $SU(5)$ or $SO(10)$ GUT}

As we have pointed out, mSUGRA makes a mild assumption that there
exists a GUT physics which encompasses MSSM beyond the unification
scale $M_{\rm G}$~\footnote{We remind the readers that inflation
occurs around a flat direction VEV $\phi_0 \sim 10^{14}$ GeV. Since
$\phi_0 \ll M_{\rm G}$, heavy GUT degrees of freedom play no role in
the dynamics of MSSM inflation, and hence they can be ignored.}. Here
we wish to understand how such embedding would affect inflationary
scenario, for instance, would it be possible to single out either
$LLe$ or $udd$ as a candidate for the MSSM inflaton.

The lowest order non-renormalizable superpotential terms which lift
$LLe$ and $udd$ are (see Eq.~(\ref{supot})):
\beq \label{supot2}
{(LLe)^2 \over M^3_{\rm P}} ~ ~ , ~ ~ {(udd)^2 \over M^3_{\rm P}}.
\eeq
It is generically believed that gravity breaks global symmetries.
Then all {\it gauge invariant} terms which are $M_{\rm P}$
suppressed should appear with $\lambda \sim {\cal O}(1)$. Obviously
the above terms in Eq.~(\ref{supot2}) are invariant under the SM.
Once the SM is embedded within a GUT at the scale $M_{\rm G}$, where
gauge couplings are unified, the gauge group will be enlarged. Then
the question arises whether such terms in Eq.~(\ref{supot2}) are
invariant under the GUT gauge group or not. Note that a GUT singlet
is also a singlet under the SM, however, the vice versa is not
correct. To answer this question, let us consider $SU(5)$ and
$SO(10)$ models separately.

\begin{itemize}

\item{{\bf $SU(5)$}:\\
We briefly recollect representations of matter fields in this case:
$L$ and $d$ belong to ${\bf {\bar 5}}$, while $e$ and $u$ belong to
${\bf 10}$ of $SU(5)$ group. Thus under $SU(5)$ the superpotential
terms in Eq.~(\ref{supot2}) read
\beq \label{su5}
{{\bf \bar {5}} \times {\bf \bar {5}}
\times {\bf 10} \times {\bf \bar {5}} \times {\bf \bar {5}} \times {\bf 10}
\over M^3_{\rm P}}.
\eeq
This product clearly includes a $SU(5)$ singlet. Therefore in the case
of $SU(5)$, we expect that $M_{\rm P}$ suppressed terms as in
Eq.~(\ref{supot}) appear with $\lambda \sim {\cal O}(1)$~\footnote{If
we were to obtain the $(LLe)^2$ term by integrating out the heavy
fields of the $SU(5)$ GUT, then $\lambda=0$. This is due to the fact
that $SU(5)$ preserves $B-L$.}.}

\item{{\bf $SO(10)$}:\\
In this case all matter fields of one generation are included in the
spinorial representation ${\bf 16}$ of $SO(10)$. Hence the
superpotential terms in Eq.~(\ref{supot2}) are $[{\bf 16}]^6$ under
$SO(10)$, which does not provide a singlet. A {\it gauge invariant}
operator will be obtained by multiplying with a $126$-plet Higgs.
This implies that in $SO(10)$ the lowest order {\it gauge invariant}
superpotential term with $6$ matter fields arises at $n=7$ level:
\beq \label{so10}
{{\bf 16} \times {\bf 16} \times {\bf 16} \times {\bf 16} \times {\bf 16}
\times {\bf 16} \times {\bf 126}_H \over M^4_{\rm P}}\,.
\eeq
Once ${\bf 126}_H$ acquires a VEV, $S0(10)$ can break down to a lower
ranked subgroup, for instance $SU(5)$. This will induce an effective
$n=6$ non-renormalizable term as in Eq.~(\ref{supot}) with
\beq \label{solam}
\lambda \sim \frac{\langle {\bf 126}_H
\rangle}{M_{\rm P}} \sim \frac{{\cal O}(M_{\rm GUT})}{M_{\rm P}}\,.
\eeq
Hence, in the case of $SO(10)$, we can expect $\lambda \sim {\cal
O}(10^{-2}- 10^{-1})$ depending on the scale where SO(10) gets
broken.}

\end{itemize}

We conclude that embedding MSSM in $SO(10)$ naturally implies
$\lambda \ll 1$. Hence an experimental confirmation of the focus
point region may be considered as an indication for $SO(10)$. More
precise determination of the spectral index $n_s$ from future
experiments (such as PLANCK) can in addition shed light on the scale
of $SO(10)$ breaking. Smaller values of $n_s$ (within the range
$0.92 \leq n_s \leq 1$) point to smaller $\lambda$, as can be seen
from  figure 6. This, according to Eq.~(\ref{solam}), implies a
scale of $SO(10)$ breaking, i.e. $\langle {\bf 126}_H \rangle$,
which is closer to the GUT scale.

Further note that embedding the MSSM within $SO(10)$ also provides
an advantage for obtaining a right handed neutrino.


\subsection{Including Right-Handed Majorana Neutrinos}

Eventually one would need to supplement MSSM with additional
ingredients to explain the tiny neutrino masses. Here we consider the
most popular framework; the see-saw mechanism which invokes MSSM plus
three RH (s)neutrinos $N_1,~N_2,~N_3$ with respective Majorana masses
$M_i$.  By adding new superfields to MSSM, one can write a larger
number of non-renormalizable gauge-invariant terms of the form in
Eq.~(\ref{supot}). As a result, a given flat direction might be lifted
at a a different superpotential level. Then a natural question arises
that whether/how adding new superfields will affect the inflaton
candidates, i.e. $LLe$ and $udd$ flat directions.

Since, $N_i$, $1 \leq i\leq 3$, are SM singlets, we can write the
following $n=4$ superpotential terms:
\beq \label{nsup}
{N_i L L e \over M_{\rm P}} ~ ~ , ~ ~ {N_i u d d \over M_{\rm P}}.
\eeq
Note that these terms are also singlet under $SU(5)$ and $SO(10)$.
In the case of $SU(5)$, the terms in Eq.~(\ref{nsup}) read ${\bf
{\bar 5}} \times {\bf {\bar 5}}\times {\bf 10} \times {\bf 1}$,
which includes a singlet. While in the case of $SO(10)$, since $N$
belongs to the ${\bf 16}$, the terms in Eq.~(\ref{nsup}) read ${\bf
16} \times {\bf 16}\times {\bf 16}\times {\bf 16}$, which includes a
singlet. Hence both terms in Eq.~(\ref{nsup}) are allowed in $SU(5)$
or $SO(10)$ embedding of MSSM as well~\footnote{In the case of
$SO(10)$ one can naturally obtain a right-handed neutrino.}.

We now analyze the case for two flat directions separately.

\begin{itemize}

\item{{\bf $LLe$}:\\
First let us consider the $LLe$ flat direction. Taking into account
of the family indices, there are $5$ independent $D$-flat directions
as such~\cite{gherghetta96}. Within MSSM, there are three directions which
are $F$-flat at the $n=3$ level, one of which survives until $n=6$.
However the term in Eq.~(\ref{nsup}) leads to three additional
$F$-term constraints $F_{N_i} = 0$, which are more than sufficient
to lift the remaining direction at the $n=4$ superpotential
level~\footnote{The gauge invariant $LLe$ direction will survive
until $n=6$ if all $M_i \gg \phi_0$.  However this is not a
phenomenologically viable situation.}.

Generically in this case we would expect $LLe$ to be lifted by a
non-renormalizable operator $n < 6$.}

\item{{\bf $udd$}:\\
Next consider the $udd$ direction. With family indices taken into
account, there are $9$ independent $D$-flat directions as
such~\cite{gherghetta96}. Within MSSM, $3$ directions are lifted by $n=4$
terms $uude/M_{\rm P}$, while the remaining $6$ will be lifted at
the $n=6$ level. Note that the superpotential term in
Eq.~(\ref{nsup}) lead to three $F$-term constraints at the $n=4$
level. Nevertheless, $3$ directions will still survive until $n=6$.}

\end{itemize}

Based on the above analysis, if we include the RH neutrinos, we
conclude that $udd$ direction is a more promising inflaton candidate
than $LLe$. The reason is that the flatness of the former will not
be lifted in the presence of physically motivated right handed
neutrino fields in addition to that of the MSSM fields.


\section{Few more examples}

Within MSSM there are other interesting possibilities of inflation
which we will discuss in this section. In the first section we will
discuss inflation with Dirac neutrinos.

\subsection{Inflation with Dirac neutrinos}

As we know by now that the inflaton potential has to be cosmologically
flat, which is suggestive of either a symmetry or a small coupling, or
both.  We will see that this property of the inflaton may be related
to the smallness of neutrino masses.  Identifying such a connection
could have important ramifications and could lead to a more
fundamental theory.  The model we will use as an example will contain
nothing but the MSSM and the right-handed neutrinos.  We will show
that a viable inflation in this model favors the correct scale for the
neutrino masses.

Let us consider the MSSM with three additional fields, namely the
right-handed (RH) neutrino supermultiplets. The relevant part of the
superpotential is

\beq \label{supot}
W = W_{\rm MSSM} + {\bf h} {\bf N} {\bf H}_u {\bf L}.
\eeq
Here ${\bf N}$, ${\bf L}$ and ${\bf H}_u$ are superfields containing
the RH neutrinos, left-handed (LH) leptons and the Higgs which gives
mass to the up-type quarks, respectively. For conciseness we have
omitted the generation indices.  We note that the RH (s)neutrinos are
singlets under the standard model (SM) gauge group. However in many
extensions of the SM they can transform non-trivially under the action
of a larger gauge group.  The simplest example is extending the SM
gauge group to $SU(3)_{\rm C} \times SU(2)_{\rm W} \times U(1)_{\rm Y}
\times U(1)_{\rm B-L}$, which is a subgroup of $SO(10)$.  Here $B$ and
$L$ denote the baryon and lepton numbers, respectively. This is the
model we consider here.  In particular, the $U(1)_{\rm B-L}$ prohibits
the RH Majorana masses~\footnote{The monomials with $B-L = 0$ will be
also $D$-flat under $U(1)_{\rm B-L}$, while those with $B-L \neq 0$
must be multiplied by an appropriate number of ${\bf N}$
superfields. In particular, ${\bf N} {\bf H}_u {\bf L}$ is now a
$D$-flat direction.}.

The active neutrino masses that arise from this are given by the usual
seesaw relation $h^2 {\langle H_u\rangle}^2/M$~\cite{seesaw,NEUT-REV},
where $\langle H_u \rangle$ is the Higgs vacuum expectation value
(VEV). Although the seesaw mechanism {\em allows} for small neutrino
masses in the presence of large Yukawa couplings, it {\em does not
require } the Yukawa couplings to be of order one.  It still allows
one to choose between the large Yukawa couplings and large Majorana
masses on the one hand, and the small Yukawas, small Majorana masses,
on the other hand.  Viable models for neutrino mass matrices have been
constructed in both limits, including the low-scale seesaw
models~\cite{deGouvea:2005er, nuMSM}, in which the Yukawa couplings
are typically of the order of
\beq \label{yukawa}
h \sim 10^{-12},
\eeq
or the same order of magnitude, as it would have in the case of Dirac
neutrinos in order to explain the mass scale $\sim {\cal O}(0.1~{\rm
eV})$ corresponding to the atmospheric neutrino oscillations detected
by Super-Kamiokande experiment.

Let us now work in the basis where neutrino masses are
diagonalized. There is a flat direction ${\bf N}_3 {\bf H}_u {\bf
L}_3$ spanned by the VEV of the lower and upper weak isospin
components of ${\bf H}_u$ and ${\bf L}_3$, respectively.
The scalar field corresponding to the flat direction is denoted by
\beq \label{flat}
\phi = {{\tilde N}_3 + H^2_u + {\tilde L}^1_3 \over \sqrt{3}}\,,
\eeq
where the superscripts refer to the weak isospin components.  One must
now include the soft SUSY breaking terms, such as the mass terms and
the $A$-term. The $A$-terms are known to play an important role in
Affleck-Dine baryogenesis~\cite{dine96}, as well as in the inflation
models based on supersymmetry~\cite{AEJM}.

The potential along the flat direction is found to be
\begin{eqnarray} \label{flatpot}
V (\phi) = \frac{m^2_{\phi}}{2} \phi^2 +
\frac{h^2_3}{12} \phi^4 \,
+ \frac{A h_3}{6\sqrt{3}} {\rm cos}
\left(\theta + \theta_h + \theta_A \right) \phi^3 \,,
\end{eqnarray}
where the flat direction mass is given in terms of the soft masses of
${\tilde N_3},~H_u$, and ${\tilde L_3}$: $m^2_{\phi} =
\left(m^2_{\tilde N_3} + m^2_{H_u} + m^2_{\tilde L_3}\right)/3$. Here
we have used the radial and angular components of the flat direction
$\phi_R + i \phi_I = \sqrt{2}\phi \, {\rm exp} \left(i \theta \right)$
, and $\theta_h,~\theta_A$ are the phases of the Yukawa coupling $h_3$
and the $A$-term, respectively. We note that the above potential does
not contain any non-renormalizable term at all.

\begin{figure}
\vspace*{-0.0cm}
\begin{center}
\epsfig{figure=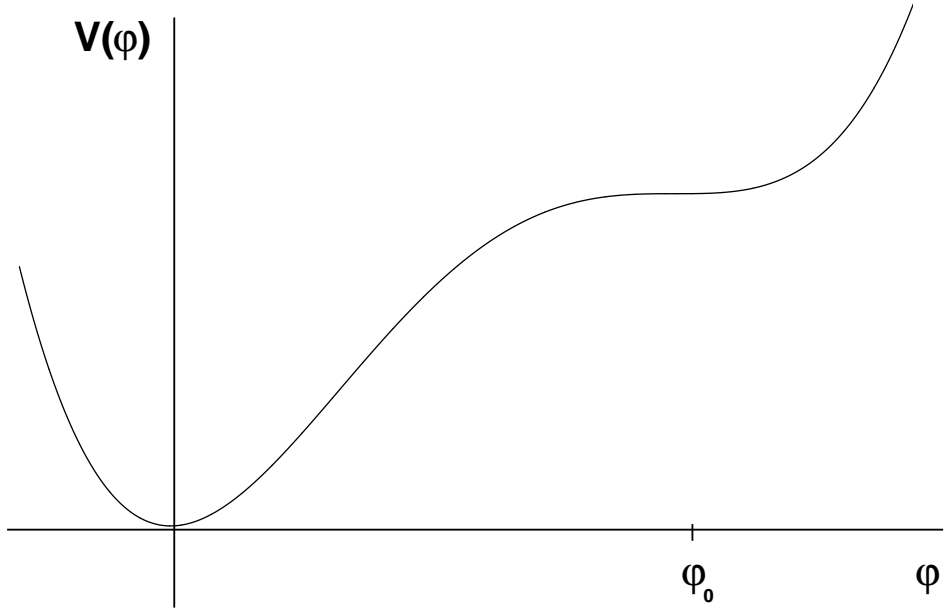,width=.84\textwidth,clip=}
\vspace*{-0.0cm}
\end{center}
\caption{ The inflaton potential. The potential is flat near the
saddle point where inflation occurs.}
\label{plot}
\end{figure}

The last term on the right-hand side of eq.~(\ref{flatpot}) is
minimized when ${\rm cos} \left(\theta + \theta_h + \theta_A \right)
= -1$. Along this direction, $V(\phi)$ has the global minimum at
$\phi=0$ and a local minimum at $\phi_0 \sim m_{\phi}/h_3$, as long
as
\beq \label{extrem} 4 m_\phi \leq A \leq 3 \sqrt{2} m_\phi \,.
\eeq
When the inequality in eq.~(\ref{extrem}) is saturated, i.e., when $A
= 4 m_{\phi}$, then both first and second derivatives of $V$ vanish at
$\phi_0$, $V^{\prime}(\phi_0)=V^{\prime\prime}(\phi_0)=0$, and the
potential becomes extremely flat in the {\it radial direction}, see
Fig.~[\ref{plot}].  We note that individually none of the terms in
eq.~(\ref{flatpot}) could have driven a successful inflation at VEVs
lower than $M_{\rm P}=2.4\times 10^{18}$~GeV. However the combined
effect of all the terms leads to a successful inflation without the
graceful exit problem.

Around $\phi_0$ the field is stuck in a plateau with potential energy
\begin{eqnarray}\label{phi0}
V(\phi_0) = {m^4_{\phi} \over 4 h^2_3} \,,~~
\phi_0 = \sqrt{3} {m_{\phi} \over h_3} \, .
\end{eqnarray}
The first and second derivatives of the potential vanish, while
the third derivative does not. Around $\phi=\phi_0$ one can expand the
potential as
$V(\phi) =V(\phi_0) + (1/ 3!)V'''(\phi_0)(\phi-\phi_0)^3$,
where
\begin{eqnarray}
\label{3rder}
V^{\prime \prime \prime}({\phi_0}) = {2 \over
\sqrt{3}} h_3 m_{\phi}\,.
\end{eqnarray}
Hence, in the range $[\phi_0 - \Delta \phi, \phi_0 + \Delta \phi]$,
where $\Delta \phi \sim H_{\rm inf}^2/V^{\prime\prime\prime}(\phi_0)
\sim \left({\phi}^3_0/M^2_{\rm P}\right) \gg H_{\rm inf}$, the
potential is flat along the real direction of the inflaton.  Inflation
occurs along this flat direction.

If the initial conditions are such that the flat direction starts in
the vicinity of $\phi_0$ with $\dot\phi\approx 0$, then a sufficiently
large number of e-foldings of inflation can be generated. Around the
saddle point, due to the random fluctuations of the massless field,
the quantum diffusion is stronger than the classical force, $H_{\rm
inf}/2\pi > \dot\phi/H_{\rm inf}$~\cite{Lindebook}, for
\beq \label{drift}
{(\phi_0-\phi) \over \phi_0} \ls \Big({m_\phi \phi_0^2 \over
M_{\rm P}^3}\Big)^{1/2}
= \left({3 m^3_{\phi} \over
h^2_3 M^3_{\rm P}}\right)^{1/2}
\, .
\eeq
At later times, the evolution is determined by the usual slow roll.
The equation of motion for the $\phi$ field in the slow-roll
approximation is $3H\dot\phi =-(1/2)V'''(\phi_0)(\phi-\phi_0)^2$.

A rough estimate of the number of e-foldings is then given by
\beq \label{efold}
{\cal N}_e(\phi) = \int {H d\phi \over \dot\phi}
\simeq \left({m_{\phi} \over 2 h_3 M_{\rm P}}\right)^2 {\phi_0 \over
(\phi_0-\phi)} ~ ,
\eeq
where we have assumed $V'(\phi) \sim (\phi - \phi_0)^2 V'''(\phi_0)$
(this is justified since $V'(\phi_0)$ and $V''(\phi_0)$ are both small). We
note that the initial displacement from $\phi_0$ cannot be much smaller than
$H_{\rm inf}$, due to the uncertainty from quantum fluctuations.

Inflation ends when the slow roll parameters become $\sim 1$. It turns
out that $\vert \eta \vert \sim 1$ gives the dominant condition
\beq \label{end} {(\phi_0-\phi) \over \phi_0}
\sim {\sqrt{3} m^3_{\phi} \over 24 h^3_3 M^3_{\rm P}}\,. \eeq
The total number of e-foldings can be computed as~\cite{AEGM,AEGJM}:
\beq \label{totalefold}
{\cal N}_{e} \sim \left({\phi_0^2 \over m_\phi
M_{\rm P}} \right)^{1/2} = \left({3 m_{\phi} \over
h^2_3 M_{\rm P}}\right)^{1/2} \,,
\eeq
evaluated after the end of diffusion, see eq.~(\ref{drift}), when the slow-roll
regime is achieved.

Let us now consider adiabatic density perturbations. As in
Ref.~\cite{AEGM,AEGJM,AKM}, one finds
\beq
\label{amp}
\delta_{H}\simeq \frac{1}{5\pi}\frac{H^2_{inf}}{\dot\phi}
\sim {h^2_3 M_{\rm P}\over 3 m_{\phi}}\,{\cal N}_{\rm COBE}^2\,.
\eeq
In the above expression we have used the slow roll approximation
$\dot\phi\simeq -V'''(\phi_0)(\phi_0- \phi)^2/3H_{\rm inf}$, and
eq.~(\ref{efold}).  The exact number depends on the scale of inflation
and on when the Universe becomes radiation dominated (we note that
full thermalization is not necessary as it is the relativistic
equation of state which matters). In our case ${\cal N}_{\rm COBE}<60$
as we shall see below.

The spectral tilt of the power spectrum and its running are
\begin{eqnarray}
\label{spect}
n_s &=& 1 + 2\eta - 6\epsilon \simeq 1 - {4\over {\cal N}_{\rm COBE}}, \\
{d\,n_s\over d\ln k} &=& - {4\over {\cal N}_{\rm COBE}^2} \,,
\end{eqnarray}
{\em cf.} \cite{AEGM}.  (We note that $\epsilon \ll 1$ while $\eta =
-2/{\cal N}_{\rm COBE}$.)

It is a remarkable feature of the model that for the weak-scale
supersymmetry and for the correct value of the Yukawa coupling,
namely,
\beq
\label{param}
m_{\phi} \simeq 100~{\rm GeV}-10~{\rm TeV}\,,~~h_3\sim 10^{-12}\,,
\eeq
the flat direction ${\bf N}_3 {\bf H}_u {\bf L}_3$ leads to a
successful low scale inflation near $\phi_0\sim
\left(10^{14}-10^{15}\right){\rm GeV} \ll M_{\rm P}$, with
\begin{eqnarray} \label{values}
&& V\sim 10^{32}-10^{36}~{\rm GeV}^4\,,~~~~H_{\inf}\sim
10~{\rm MeV}-1{\rm GeV}\,, \nonumber \\
&& {\cal N}_e\sim 10^{3}\,,~~~~T_{max}\sim 10^{8}-10^{9}~{\rm GeV}\,.
\end{eqnarray}
The total number of e-foldings driven by the slow roll inflation,
${\cal N}_e \sim 10^3$, is more than sufficient to produce a patch of
the Universe with no dangerous relics.  Those domains that are
initially closer to $\phi_0$ enter self-reproduction in eternal
inflation.  Since the inflaton, ${\bf N}_3 {\bf H}_u {\bf L}_3$,
couples directly to MSSM particles, after inflation the field
oscillates and decays to relativistic MSSM degrees of freedom.  The
highest temperature during reheating is $T_{max}\sim V^{1/4}$.  This
temperature determines the total number of e-foldings required for the
relevant perturbations to leave the Hubble radius during inflation; in
our case it is roughly ${\cal N}_{\rm COBE}\sim 50$.

Despite the low scale of inflation, the flat direction can generate
density perturbations of the correct size for the parameters listed
above.  Indeed, from eqs.~(\ref{amp},\ref{spect}), and (\ref{param}),
we obtain: $\delta_{H}\sim 10^{-5}$. Following the discussion of
Section 4.4, we obtain $0.91 \leq~n_{s}\leq 1.0$ with a negligible
running.  The spectral tilt agrees with the current WMAP 3-years' data
within $2\sigma$~\cite{WMAP3}. The tensor modes are negligible because
of the low scale of inflation.

We emphasize that the VEV of the flat direction is related to the Yukawa
coupling that can generate the Dirac neutrino mass $\sim 0.1$~eV.
The scale of the neutrino mass appears to be just right to get the
correct amplitude in the CMB perturbations.

The inflaton has gauge couplings to the electroweak and $U(1)_{\rm
B-L}$ gauge/gaugino fields. It therefore induces a VEV-dependent
mass $\sim g \langle \phi \rangle$ for these fields ($g$ denotes a
typical gauge coupling). After the end of inflation, $\phi$ starts
oscillating around the global minimum at the origin with a frequency
$m_{\phi} \sim 10^3 H_{\rm inf}$, see eq.~(\ref{values}). When the
inflaton passes through the minimum, $\langle \phi\rangle = 0$, the
induced mass undergoes non-adiabatic time variation. This results in
non-perturbative particle production~\cite{PREHEAT}.


\subsection{Inflation in gauge mediated scenarios}

In a Gauge Mediated Supersymmetry Breaking (GMSB) scenario the
two-loop correction to the flat direction potential results in a
logarithmic term above the messenger scale, i.e. $\phi >
M_S$~\cite{dGMM}. Together with the $A$-term this leads to the scalar
potential
\beq \label{scpot}
V = M_{F}^4\ln\left(\frac{\phi^2}{M_{S}^2}\right) + 
A\cos(n \theta  + \theta_A)
\frac{\lambda_{n}\phi^n}{n\,M^{n-3}_{\rm GUT}} + \lambda^2_n 
\frac{{\phi}^{2(n-1)}}{M^{2(n-3)}_{\rm GUT}}\,,
\eeq
where $M_F \sim (m_{SUSY} \times M_S)^{1/2}$ and $m_{SUSY} \sim 1$~TeV
is the soft SUSY breaking mass at the weak scale. For $\phi >
M^2_F/m_{3/2}$, usually the gravity mediated contribution, $m^2_{3/2}
\phi^2$, dominates the potential where $m_{3/2}$ is the gravitino
mass.  Here we will concentrate on the VEVs $M_s \ll \phi \ls
M^2_F/m_{3/2}$.

Although individual terms are unable to support a sub-Planckian VEV
inflation, but as shown in Refs.~\cite{AEGM,AKM,AEGJM,AM}, a
successful inflation can be obtained near the saddle point, which we
find by solving, $V^{\prime}(\phi_0) = V^{\prime\prime}(\phi_0)=0$
(where derivative is w.r.t $\phi$).
\begin{eqnarray}
\label{saddle2}
\phi_0 &=& \left( \frac{M^{n-3}_{\rm GUT} M_F^2}{\lambda_n}
\sqrt{\frac{n}{(n-1)(n-2)}} \right)^{1/(n-1)}\,, \\ 
A &=& \frac{4(n-1)^2
\lambda_n}{n M^{n-3}_{\rm GUT}} \phi_0^{n-2}\,. \label{A2}
\end{eqnarray}
In the vicinity of the saddle point, we obtain the total energy
density and the third derivative of the potential to be~\cite{AJM}:
\begin{eqnarray}
\label{V0}
V(\phi_0) &=& M_F^4 \left[ \ln\left(\frac{\phi_0^2}{M_S^2} \right) -
  \frac{3n-2}{n(n-1)} \right]\,, \\
V^{\prime\prime\prime}(\phi_0)& =& 4n(n-1) M_F^4 \phi_0^{-3}\,.
\end{eqnarray}
There are couple of interesting points, first of all note that the
scale of inflation is extremely low in our case, barring some small
coefficients of order one, the Hubble scale during inflation is given
by:
\beq
\label{Hinf}
H_{\rm inf}\sim {M_{F}^2}/{M_{\rm P}} \sim 10^{-3}-10^{-1}~{\rm eV} \,,
\eeq
for $M_F \sim 1-10$ TeV. For such a low scale inflation usually it is
extremely hard to obtain the right phenomenology. But there are
obvious advantages of having a low scale inflation, $M_F \gg H_{\rm
inf}$. The supergravity corrections and the Trans-Planckian
corrections are all negligible~\cite{AEGJM}, therefore the model
predictions are trustworthy.

Perturbations which are relevant for the COBE normalization are
generated a number ${\cal N}_{\rm COBE}$ e-foldings before the end of
inflation. The value of ${\cal N}_{\rm COBE}$ depends on thermal
history of the universe and the total energy density stored in the
inflaton, which in our case is bounded by, $V_0 \leq 10^{16}~({\rm
GeV})^4$. The required number of e-foldings yields in our case, ${\cal
N}_{\rm COBE}\sim 40$~\cite{BURGESS-multi}, provided the universe
thermalizes within one Hubble time. Although within SUSY
thermalization time scale is typically very long~\cite{AVERDI2},
however, in this particular case it is possible to obtain a rapid
thermalization.

Near the vicinity of the saddle point, $\phi_0$, the potential is
extremely flat and one enters a regime of
self-reproduction~\cite{Lindebook}. The self-reproduction regime lasts as
long as the quantum diffusion is stronger than the classical drag;
$H_{inf}/2\pi >\dot\phi/H_{inf}$, for $\phi_s \leq \phi \leq \phi_0$,
where $\phi_0 - \phi_s \simeq M_{F}\left({\phi_0}/{M_{\rm
P}}\right)^{3/2}$.  From then on, the evolution is governed by the
classical slow roll. Inflation ends when $\epsilon \sim 1$,
which happens at $\phi \simeq \phi_e$, where~\cite{AJM}
\beq
\phi_e-\phi_0=-\sqrt{\frac{V_0\phi_0^3}{\sqrt{2}n(n-1)M_F^4M_{\rm P}}}\,.
\eeq
Assuming that the classical motion is due to the third derivative of
the potential, $V^{\prime}(\phi)\simeq (1/2)
V^{\prime\prime\prime}(\phi_0)(\phi-\phi_0)^2$, the total number of
e-foldings during the slow roll period is found to be~\cite{AJM}:
\beq
\label{N}
{\cal N}_{tot}=\int\frac{H d\phi}{\dot\phi}\simeq
\frac{2V_0\phi_0^3}{4n(n-1)M_{F}^4M_{\rm P}^2}\Big(\frac{1}{\phi_0-\phi_s}\Big)
\,.
\eeq
This simplifies to~\cite{AJM}
\beq
\label{N1}
{\cal N}_{\rm tot}\simeq {\phi_{0}^{3/2}}/({M_{\rm P}^{1/2}M_F})\,.
\eeq
Let us now consider the adiabatic density perturbations. Despite
$H_{\rm inf}\ll 1$~eV, the flat direction can generate adequate
density perturbations as required to explain to match the
observations. Recall that inflation is driven by $V^{\prime \prime
\prime}\neq 0$, we obtain~\cite{AJM}
\beq
\label{amp}
\delta_{H}\simeq ({1/5\pi})({H^2_{inf}}/{\dot\phi})
\sim {M_F^2 M_{\rm P} {\cal N}_e^2}{\phi_0^{-3}}\sim 10^{-5}\,.
\eeq
Note that for $M_F \sim 10$~TeV, and ${\cal N}_{\rm COBE}\sim 40$, we
match the current observations~\cite{WMAP3}, when $\phi_0\sim
10^{11}$~GeV. The validity of Eq.~(\ref{scpot}) for such a large VEV
requires that $M^2_F > (10^{11}~{\rm GeV}) \times m_{3/2}$. For $M_F
\sim 10$ TeV this yields the bound on the gravitino mass, $m_{3/2} <
1$ MeV, which is compatible with the dark matter constraints as we
will see.

We can naturally satisfy Eq.~(\ref{amp}) provided, $n=6$. The
non-renormalizable operator, $n=6$, points towards two MSSM flat
directions out of many,
\beq
LLe ~~~~~{\rm and}~~~~~ udd\,.
\eeq
As we discussed before in \cite{AEGM}, these are the only directions
which are suitable for inflation as they give rise to a non-vanishing
$A$-term. Note that the inflatons are now the gauge invariant objects.
The total number of e-foldings, during the slow roll inflation, after
using Eq.~(\ref{N1}) yields,
\beq
\label{Ntot}
{\cal N}_{\rm tot}\sim 10^{3}\,.
\eeq
While the spectral tilt and the running of the power spectrum are
determined by ${\cal N}_{\rm COBE} \sim 40 \ll {\cal N}_{\rm tot}$.
\begin{eqnarray}
\label{spect}
& & n_s = 1 + 2\eta - 6\epsilon \simeq 1 - \frac{4}{{\cal N}_{\rm COBE}} 
\sim 0.90\,,\\
\label{runspect}
& & \frac{d\,n_s}{d\ln k} = 16\epsilon\eta - 24 
\epsilon^2 - 2\xi^2 \simeq - \frac{4}{{\cal N}_{\rm COBE}^2} 
\sim - 10^{-3} \,, \nonumber \\
& & \,
\end{eqnarray}
where $\xi^2 = M_P^4 V^{\prime} V^{\prime\prime\prime}/ V^2$. Note
that the spectral tilt is slightly away from the $2\sigma$ result of
the current WMAP 3 years data, on the other hand running of the
spectrum is well inside the current bounds~\cite{WMAP3}.

At first instance one would discard the model just from the slight
mismatch in the spectral tilt from the current observations. However
note that our analysis strictly assumes that the slow roll inflation
is driven by $V^{\prime\prime\prime}(\phi_0)$. This is particularly
correct if $V^{\prime}(\phi_0) =0$ and
$V^{\prime\prime}(\phi_0)=0$. Let us then study the case when
$V^{\prime}(\phi_0)\neq 0$, as discussed in~\cite{AEGJM,AM}.

The latter case can be studied by parameterizing a small deviation
from the exact saddle point condition by solving near the point of
inflection, where we wish to solve $V^{\prime\prime}(\phi_0)=0$ and we
get up-to 1st order in the deviation, $\delta<1$,
\beq
\label{inflection1}
\tilde{A} = A(1-\delta), \qquad \tilde{\phi}_0 = \phi_0 \left( 1 -
  \frac{n-1}{n(n-2)} \delta \right)\,,
\eeq
with $A$ and $\phi_0$ are the saddle point solutions. Then the 1st
derivative is given by
\beq
\label{a1}
V^{\prime}(\phi_0) = 4 \frac{n-1}{n-2} M_F^4 \phi_0^{-1} \delta\,.
\eeq
Therefore the slope of the potential is determined by,
$V^{\prime}(\phi) \simeq V^{\prime}(\phi_0) + (1/2) V^{\prime \prime
\prime}(\phi_0) (\phi - \phi_0)^2$.

Note that both the terms on the right-hand side are positive.  The
fact that $V^{\prime}(\phi_0) \neq 0$ can lead to an interesting
changes from the saddle point behavior, for instance the total number
of e-foldings is now given by
\beq 
\label{tot2}
{\cal N}_{\rm tot} = \frac{V(\phi_0)}{M^2_{\rm P}}
\int_{\phi_{\rm end}}^{\phi_0} \frac{d\phi}{{V^{\prime}(\phi_0) + 
\frac{1}{2} V^{\prime \prime \prime}(\phi_0) (\phi - \phi_0)^2}}\,.
\eeq
First of all note that by including $V^{\prime}$, we are slightly away
from the saddle point and rather close to the point of inflection.
This affects the total number of e-foldings during the slow roll. It
is now much less than that of ${\cal N}_{\rm tot}$, i.e. ${\cal
N}_{\rm tot}\ll 10^3$, see Eq.~(\ref{Ntot}).

When both the terms in the denominator of the integrand contributes
equally then there exists an interesting window.
\beq\label{constr007}
{\kappa}/{8}\leq \delta \leq {\kappa}/{2}\,.
\eeq
where
\beq\label{kappa} 
\kappa \equiv \frac{n-2}{n(n-1)^2} \left[ \ln\left(
\frac{\phi_0^2}{M_S^2} \right) - \frac{3n-2}{n(n-1)} \right]^2 
\frac{\phi_0^4}{M_{\rm P}^4 {\cal N}_{\rm COBE}^2}\,.
\eeq
The lower limit in Eq.~(\ref{constr007}) is saturated when
$V^{\prime}(\phi_0)=0$, while the upper limit is saturated when ${\cal
N}_{\rm tot}\simeq {\cal N}_{\rm COBE} \simeq 40$. It is also easy to
check that there will be no self-reproduction regime for the field
values determined by $\delta$.

It is a straightforward but a tedious exercise to demonstrate that
when the upper limit of Eq.~(\ref{constr007}) is saturated the
spectral tilt becomes $n_s\simeq 1$, when the lower limit is satisfied
we recover the previous result with $n_s=0.9$. This value, $n_s
\rightarrow 1$, can be easily understood as $\phi_{\rm COBE}
\rightarrow \phi_0$ (where $\phi_{\rm COBE}$ corresponds to the VeV
where the end of inflation corresponds to ${\cal N}_{\rm COBE}\sim
40$), in which case, $\eta \rightarrow 0$.  Therefore the spectral
tilt becomes nearly scale invariant. We therefore find a
range~\cite{AEGJM,AM,AJM},
\beq
0.90 \leq n_s\leq 1\,,
\eeq
whose width is within the $2\sigma$ error of the central
limit~\cite{WMAP3}.  Similarly the running of the spectral tilt gets
modified too but remains within the observable limit~\footnote{A
similar exercise can be done for the running of the spectral tilt and
the running lies between $-16/{\cal N}_{\rm COBE}^2 \leq d n_s/d\ln
k\leq -4/{\cal N}_{\rm COBE}^2$~\cite{AEGJM,AM}.}, while the amplitude
of the power spectrum is least affected~\cite{AEGJM,AM,AJM,ADM}.

Let us now discuss the issue of reheating and thermalization.
Important point is to realize that the inflaton belongs to the MSSM,
i.e. $LLe$ and $udd$, both carry MSSM charges and both have gauge
couplings to gauge bosons and gauginos. After inflation the condensate
starts oscillating.  The effective frequency of the inflaton
oscillations in the Logarithmic potential, Eq.~(\ref{scpot}), is of
the order of $M^2_{F}/\phi_0$, while the expansion rate is given by
$H_{\rm inf} \sim M^2_F/M_{\rm P}$. This means that within one Hubble
time the inflaton oscillates nearly $M_{\rm P}/\phi_0 \sim 10^{7}$
times. The motion of the inflaton is {\it strictly } one dimensional
from the very beginning. During inflation, the imaginary direction is
very heavy and settles down in the minimum of the potential. This also
prohibits fragmentation of the flat direction to form
Q-balls~\cite{MSSM-REV,AEGJM}. Although one could still argue positive
and negative charged Q-ball formation~\cite{SURFACE}, but they do not
form adequately to alter reheating and thermalization in any
significant way.

An efficient bout of particle creation occurs when the inflaton
crosses the origin, which happens twice in every oscillation. The
reason is that the fields which are coupled to the inflaton are
massless near the point of enhanced symmetry. Mainly electroweak gauge
fields and gauginos are then created as they have the largest coupling
to the flat direction.  The production takes place in a short
interval. Once the inflaton has passed by the origin, the gauge
bosons/gauginos become heavy by virtue of VeV dependent masses and
they eventually decay into particles sparticles, which creates the
relativistic thermal bath. This is so-called instant preheating
mechanism~\cite{INSTANT}.  In a favorable condition, the flat
direction VeV coupled very weakly to the flat direction inflaton could
also enhance the perturbative decay rate of the inflaton~\cite{ABM}.
In any case there is no non-thermal gravitino production~\cite{MAROTO}
as the energy density stored in the inflaton oscillations is too low.

A full thermal equilibrium is reached when ${\it a)~kinetic }$ and
${\it b)~chemical~equilibrium }$ are established~\cite{AVERDI1}. The
maximum temperature of the plasma is given by
\beq
\label{tmax}
T_{\rm R} \sim [V(\phi_0)]^{1/4} \sim M_F \ls 10 {\rm
TeV}\,, 
\eeq
when the flat direction, either $LLe$ or $udd$ evaporates completely.
This naturally happens at the weak scale. There are two very important
consequences which we summarize below.

\subsubsection{Cold electroweak Baryogenesis}

The model strongly points towards electroweak baryogenesis within
MSSM.  Note that the reheat temperature is sufficient enough for a
thermal electroweak baryogenesis~\cite{BARYO-REV}.

However, if the thermal electroweak baryogenesis is not triggered,
then cold electroweak baryogenesis is still an option~\cite{GKS}.
During the cold electroweak baryogenesis, the large gauge field
fluctuations give rise to a non-thermal sphaleron transition.  In our
case it is possible to excite the gauge fields of $SU(2)_{L}\times
U(1)_{Y}$ during instant preheating provided the inflaton is $LLe$.
The $LLe$ as an inflaton carries the same quantum number which has a
$B-L$ anomaly and large gauge field excitations can lead to
non-thermal sphaleron transition to facilitate baryogenesis within
MSSM.


\subsubsection{Gravitino dark matter}

Within GMSB gravitinos are the LSP and if the $R$-parity is conserved
then they are an excellent candidate for the dark matter. There are
various sources of gravitino production in the early
universe~\cite{STEFFAN,AHJMP}. However in our case the thermal
production is the dominant one and mainly helicity $\pm 1/2$
gravitinos are created. Gravitinos thus produced have the correct dark
matter abundance for~\cite{dGMM,Buchmuller,AJM}
\beq \label{dm}
{m_{3/2} \over 100~ {\rm KeV}} \simeq {1 \over {\rm few}} \Big({T_{\rm R} 
\over 1~{\rm TeV}}\Big) \Big({M_{\tilde g} \over 1~{\rm TeV}}\Big)^2,
\eeq
where $M_{\tilde g}$ is the gluino mass. For $m_{3/2} \gs 100$ KeV, 
Eq.~(\ref{dm}) is easily satisfied for $M_{\tilde g} \sim 1$ TeV and
$T_{\rm R} \ls 10$ TeV. We remind that for ${\cal
O}({\rm keV}) \ls m_{3/2} < 100$ KeV gravitinos produced from the
sfermion decays overclose the universe~\cite{dGMM}.

\section{Quantum initial conditions, cosmological constant and 
a string landscape}

One important result is that the MSSM inflation produces many
e-foldings of slow roll inflation, $\mathcal{N}_{e}\sim 10^{3}$, with
a preceding self-reproduction regime.  One difficulty of MSSM
inflation, though, is that it requires a fine-tuning of the initial
conditions for slow-roll~\cite{AEGJM}.

Our aim is to address this initial condition issue by studying the
quantum fluctuations of the MSSM inflaton during a false vacuum
inflation which naturally fine tunes the initial conditions for MSSM
inflation, i.e. why the MSSM inflation occurs near the saddle point of
the potential.  The idea is to have an early bout of false vacuum
during which the MSSM inflaton will obtain false vacuum induced
quantum fluctuations from the point of enhanced symmetry to the saddle
point. Once the false vacuum tunnels to a lower cosmological constant
or the vacuum energy density comparable to that of the MSSM inflation
then the latter potential takes over for a last stage of MSSM driven
inflation.

Without a prior phase of inflation it would be difficult to explain
the initial condition otherwise. For example, one may not invoke
thermal effects to trap the flat direction near the saddle point as it
does not correspond to a false minimum, or a point of enhanced
symmetry.  As we shall see, old inflation and MSSM inflation
complement each other nicely.

Also, there is some evidence that string theory has a ``landscape'' of
metastable vacua with varying cosmological constant, moduli VEVs (and
therefore couplings), supersymmetry (SUSY) breaking scale, and so on,
which can be studied using statistical arguments (see
\cite{hep-th/0409207,hep-th/0601053,Kachru} for reviews).  Since many
of these metastable vacua have large energy densities (as we describe
below), it seems likely that the early universe could have existed as
a de Sitter spacetime with a large cosmological constant, which then
decayed by tunnelling, as in old inflation \cite{Guth}.  In fact, such
a cosmology with multiple stages of inflation \cite{BURGESS-multi}
provides a mechanism by which the full landscape of vacua is
populated, as in \cite{Abbott:1984qf,hep-th/0004134,hep-th/0408133}.
With some caveats, this mechanism can also relax the cosmological
constant quickly
\cite{Abbott:1984qf,hep-th/0004134,hep-th/0408133,hep-th/0611148,
hep-th/0612056}.  In particular, the cosmological constant must be
able to decay even though other sectors will dominate once the energy
density reaches $\sim 1\, (\textnormal{TeV})^4$.

One obvious worry about this picture is that, if the universe were to
tunnel out of the false vacuum, then the universe would be devoid of
any entropy as the nucleated bubble would keep expanding forever with
a negative spatial curvature.  Such a universe would have no place in
a real world, so it is important that the last stage of inflation be
driven in an observable sector, such as in the case of MSSM inflation
which not only produces the right amplitude of density spectrum but
also provides us with a desired thermal entropy,and cold dark matter
abundance.


\subsection{Old Inflation on the Landscape}\label{s:landscape}

The most basic fact about the landscape (which is usually not
discussed, being of little interest for present-day physics) is that
the cosmological constant is generically large.  A simple way to see
that follows \cite{hep-th/0004134}, which describes the landscape
contribution to the cosmological constant as arising from string
theory flux.  In this picture, the vacuum energy
\be\lb{bp}
V=M_P^2\Lambda = M_P^2\Lambda_0 +\alp{}^{-2}\sum_i c_i n_i^2\ ,
\ee
where $c_i\lesssim 1$ are constants and $n_i$ are flux quantum numbers
(note that $\Lambda$ has dimension $\mathrm{mass}^2$ in our notation
as it enters the Einstein equation as $\Lambda g_{\mu\nu}$).  It is
clear that large $\Lambda$ corresponds to a large radius shell in the
space of flux quanta, so larger $\Lambda$ will have more possible
states.  All in all, string theory (from the landscape point of view)
could have from $10^{500}$ to even $10^{1000}$ vacua
\cite{hep-th/0409207,hep-th/0601053,Kachru,hep-th/0004134}, with the
vast majority of those having large cosmological constants.


\subsection{Decay time scales and inflation}\label{s:decaytime}

The decay rate per volume (by tunnelling) of a metastable vacuum to
the nearest neighboring vacuum\footnote{To avoid subtleties, we
require that the both states have nonnegative energy density.  See,
for example, \cite{hep-th/0211160} for issues in the negative
$\Lambda$ case.  Also, \cite{hep-th/0701083} has discussed the
importance of negative $\Lambda$ vacua in possibly separating parts of
the landscape from each other.  We adopt the view, as discussed in
that paper, that the landscape is sufficiently complicated that there
are no isolated regions.} takes the form
\be\lb{tunnel}
\Gamma/V = C \exp\left(-\Delta S_E\right)\ ,\ee
where $C$ is a one-loop determinant and $\Delta S_E$ is the difference
in Euclidean actions between the instanton and the background with
larger cosmological constant.  The determinant $C$ can at most be
$C\lesssim M_P^4$, simply because $M_P$ is the largest scale
available, and estimates (ignoring metric fluctuations) give a value
as small as $C\sim r^{-4}$, with $r$ the instanton bubble radius
\cite{hep-ph/9308280}.  If we therefore look at a decay rate in a
(comoving) Hubble volume, we find
\be\lb{tunnel2}
\Gamma\lesssim \frac{M_P^4}{H^3}\exp\left(-\Delta S_E\right)\,.
\ee
Especially with a large Hubble scale, the associated decay time is
much longer than $1/H$, given that typically $\Delta S_E\gg 1$.

In fact, as given in \cite{hep-th/0305018} following
\cite{Coleman:1980aw,Brown:1987dd,Brown:1988kg}, the Euclidean action
takes the form
\be\lb{action}
\Delta S_E = 2\pi^2 r^3 \tau_e \,,
\ee
where the bubble radius and effective tension $\tau_e$ are given in
terms of dimensionless cosmological constants
$\lambda=M_P^4\Lambda/\tau^2$, and the actual tension of the bubble
wall $\tau$.  The full formulae are listed in the appendix.

Unfortunately, a given tunnelling process in the landscape probably
has $\lambda_+\sim 1$ and $\lambda_-\lesssim 1$.  However, it may be
instructive to look at three limits, $\lambda_+\to 0,\infty$.  In the
latter case, we simultaneously take $\lambda_-\to 0$ or
$\lambda_-\to\lambda_+$ (these are the same for $\lambda_+\to 0$).  As
$\lambda_+$ vanishes, we find
\be\lb{mink}
\Delta S_E\to 24\pi^2 \frac{M_P^6}{\tau^2\lambda_+}=24\pi^2
\frac{M_P^2}{\Lambda_+}\, .
\ee
(The tension surprisingly drops out!)  For $\lambda_+\to\infty$, we
find the limits
\bea
\Delta S_E&\to& 12\pi^2\frac{M_P^6}{\tau^2\lambda_+}=
12\pi^2\frac{M_P^2}{\Lambda_+}\ \ (\lambda_-\to 0)\ ,\lb{drop}\\
\Delta S_E&\to& 6\sqrt{3}\pi^2\frac{M_P^6}{\tau^2\lambda_+^{3/2}}
=6\sqrt{3}\pi^2\frac{\tau}{\Lambda_+^{3/2}}\ \ (\lambda_-\to\lambda_+)\, .
\lb{step}
\eea
The only parametrically different behavior is the last limit as
$\lambda_{\pm}\to\infty$, which means either that the tension is very
small or that the cosmological constants are both very large (or that
$M_P\to\infty$, corresponding to field theory without gravity).
Technically, the instanton approximation can break down in the limit
(\ref{step}), if the Euclidean action becomes much smaller than one
(as a full quantum mechanical treatment would be necessary), but the
important point is that $\Gamma/H\gtrsim 1$ if that is the case.  From
(\ref{tunnel2}), we might expect that $\Gamma/H\gg 1$, but we should
remember that the coefficient of the exponential in (\ref{tunnel2}) is
an upper limit and that quantum mechanical effects could limit the
decay rate to $\Gamma \sim H$.  In section \ref{s:mssminflate}, we
give an improved form of the argument of \cite{hep-th/0004134} that
shows how old inflation on the landscape can source MSSM inflation
without inflaton trapping and with only small jumps in the
cosmological constant.

Note that, when MSSM inflation starts, the ``bare'' cosmological
constant (that not associated with the MSSM inflaton) might still be
considerably larger than the present value.  This means that further
instanton decays should take place to reduce the bare cosmological
constant, and these decays should occur during MSSM inflation in order
for the bubble regions to grow long enough.  Since it seems likely
that the instanton bubble tension will be large compared to the scale
of MSSM inflation, the decay rate will be given by (\ref{mink}), which
is highly suppressed.  This would then require MSSM inflation to last
for extremely many e-foldings.  Fortunately, MSSM inflation naturally
includes a self-reproduction (eternal inflation) regime prior to
slow-roll \cite{AEGM,AEGJM}.  In other words, it seems that a
low-scale eternal inflation is a necessary part of using decays to
solve the cosmological constant problem, and MSSM inflation includes
it naturally.  Eternal inflation solves several of problems relating
false vacuum inflation to observations, and it is particularly viable
at the TeV scale of MSSM inflation.


\subsection{Initial quantum kicks for an MSSM fields}\label{s:mssminflate}

Let us now discuss the what happens to the observable sector in the
background of inflation spacetime. As we have noticed, the universe
cascades from large cosmological constant to another somewhat smaller
as time progresses.  Given the decay time, it is fairly evident that
huge number of e-foldings can be generated.

During the false vacuum inflation the energy density of the universe,
and hence the expansion rate $H_{\mathrm{false}}$, remains constant
for a given $\Lambda$.

The flat direction potential receives corrections from soft SUSY
breaking mass term, the non-renormalizable superpotential correction,
and corrections due to the large Hubble expansion for a minimal choice
of K\"ahler potential. The Hubble correction is relevant only when
$m_{\phi}\ll H(t)$, which is parameterized by two constants $c\sim
\mathcal{O}(1)$ and $a\sim \mathcal{O}(1)$, result in
\cite{dine95,dine96,MSSM-REV,Kusenko}\footnote{The origin of the
Hubble induced terms is due to couplings between the modulus which
drives the false vacuum inflation and the MSSM sector. Apriori, even
in string theory, the K\"ahler potential for the modulus is not well
known at all points of the parameter space. Similarly, within the
MSSM, the K\"ahler potential is unknown. Therefore, the coefficients
$c,a$ are not fixed. For a no-scale type model, the Hubble induced
corrections are vanishing at tree level; however, they do appear at
one loop $| c | \sim 10^{-2}$ \cite{GMO,ADM1}.}
\be \label{hubblepot} V =
\frac{1}{2} (m^2_{\phi} + c H^2_{\mathrm{false}}) | \phi |^2 +
\left[(A + a H_{\mathrm{false}}) \lambda_n \frac{\phi^n}{n
M^{n-3}_{P}} + {\mathrm{h.c.}})\right] + \lambda^2_n \frac{|\phi
|^{2(n-1)}} {M^{2(n-3)}_{P}}\,, \ee 
where the soft SUSY breaking mass term is generically small compared
to the Hubble expansion rate of the false vacuum, $m_{\phi}\sim 1\
\mathrm{ TeV}\ll H_{\mathrm{false}}$. We define $\Phi\equiv \phi
e^{in\theta}$, and the dynamics of $\phi$ in an inflationary
background depends on $c$ and $a$.  Therefore we consider different
cases separately.

It turns out that the dynamics of the MSSM flat direction largely
follow the physics discussed in \cite{hep-th/0004134}.  We review and
elaborate on their discussion in the context of the MSSM and give an
important improvement on their result in the context of negligible
Hubble corrections.


\subsubsection{Positive Hubble induced corrections}

A positive Hubble induced correction provides $c\sim +\mathcal{
O}(1),a\sim \mathcal{O}(1)$. This is a typical scenario when the
K\"ahler potential for the string modulus comes with a canonical
kinetic term. Although this could be treated as a special point on the
K\"ahler manifold, it is nevertheless important to discuss this
situation. This is also the simplest scenario out of all
possibilities. A generic flat direction gets a large mass of the order
of Hubble expansion rate; therefore, its fluctuations are unable to
displace the flat direction from its global minimum.

On phenomenological grounds, this is an uninteresting and undesirable
case, since the bubble does not excite any of the MSSM
fields. Therefore, the bubble remains empty and devoid of energy with
no graceful exit of inflation from the false vacuum. The universe
continues cascading to smaller $\Lambda$ with smaller
$H_{\mathrm{false}}$.  Eventually, when $H_{\mathrm{false}}\leq
m_{\phi}$, the MSSM flat directions would be free to move. However,
through this time the fields were never displaced from their minimum
and therefore the dynamics of the flat directions would remain
frozen. The universe would be cold, as it was before, and the spatial
curvature would remain negative. It is fair to say that, on
phenomenological grounds, such a universe is already ruled out.  This
paves the way for more interesting scenarios, which we discuss next.


\subsubsection{Negligible Hubble induced corrections}

In this case, the potential is not affected by the false vacuum
inflation at all, namely $|c|,|a|\ll 1$.  So long as $V^{\prime
\prime}(\phi) \ll H^2_{\mathrm{false}}$, the flat direction field
$\phi$ makes a quantum jump of length $H_{\mathrm{false}}/2\pi$ within
each Hubble time.\footnote{To be more precise, the quantum
fluctuations of $\phi$ have a Gaussian distribution, and the r.m.s.~of
modes which exit the horizon within one Hubble time is
$H_{\mathrm{false}}/2 \pi$.} These jumps superpose in random walk
fashion resulting in \cite{Lindebook}
\be \label{diff}
\Big(\frac{d \langle \phi^2 \rangle}{dt}\Big)_{fluctuations} =
\frac{H^3_{\mathrm{false}}}{4 \pi^2}.
\ee
On the other hand, the classical slow roll due to the potential leads to
\be \label{slow}
\Big(\frac{d \langle \phi^2 \rangle}{dt}\Big)_{slow~roll} = - \frac
{2 \langle V^{\prime}(\phi) \phi \rangle}{3 H_{\mathrm{false}}}.
\ee
For a massive scalar field $V(\phi)\sim m^2_{\phi} \phi^2/2$, the
combined effects yield \cite{Lindebook}
\be \label{fluct}
\langle \phi^2 \rangle = \frac{3 H^4_{\mathrm{false}}}{8 \pi^2 m^2_{\phi}}
\left[1 - \exp\left(-\frac{2 m^2_{\phi}}{3 H_{\mathrm{false}}}
t\right)\right]\,.
\ee
The maximum field value
\be \label{rms}
\phi_{r.m.s.} = \sqrt{\frac{3}{8 \pi^2}} \frac{H^2_{\mathrm{false}}}{m_{\phi}},
\ee
at which the slow roll motion (\ref{slow}) counterbalances the random
walk motion (\ref{diff}), is reached for $\Delta t \gg
3H_{\mathrm{false}}/2 m^2_{\phi}$. This amounts to a number
\be
\label{falsefold}
\mathcal{N}_{\mathrm{false}} \gg \frac{3}{2}
\left(\frac{H_{\mathrm{false}}}{m_{\phi}}\right)^2.
\ee
of e-foldings of inflation in the false vacuum.

In the absence of Hubble induced corrections, the potential
in~(\ref{hubblepot}) has a saddle point at
\be \label{saddle}
\phi_{0}=\left(\frac{m_{\phi}M_{P}^{n-3}}
{\lambda_{n}\sqrt{2n-2}}\right)^{1/(n-2)}\,,
\ee
where $V^{\prime}(\phi_0)=V^{\prime\prime}(\phi_0)=0$ while
$V^{\prime\prime\prime}(\phi_0)\neq 0$, provided that
$A^2=8(n-1)m_{\phi}^2$.\footnote{This can happen in the gravity
mediated case, where $A \sim m_{\phi} \sim m_{3/2}$, where
$m_{3/2}\sim \mathcal{O}(1~\mathrm{TeV})$ is the gravitino mass. The
situation is quite different in the case of a gauge mediated SUSY
breaking scenario \cite{AJM}.}  For $m_{\phi}\sim 100~\mathrm{GeV}-10$
TeV and $\lambda_n \sim {\cal O}(1)$, and for $n=6$, the VEV is
$\phi_0\sim 10^{14}-10^{15}$ GeV. The suitable flat directions are
$LLe$ and $udd$, which are lifted by $n=6$ superpotential terms and
also have a non-zero $A$-term as required by the condition for a
saddle point \cite{AEGM,AEGJM}.\footnote{In order to have successful
inflation, we require the condition $A^2=8(n-1)m_{\phi}^2$ to be
satisfied to one part in $10^{9}$. Although this requires a fine
tuning, SUSY can allow it to be maintained order by order if
$A/m_{\phi}$ acts as an infrared fixed point of the renormalization
group flow \cite{ADM1}. Also, as noted in Ref.~\cite{AFM}, this tuning
can be explained naturally by the landscape picture.  For larger
deviations there is a point of inflection with large
$V^{\prime}(\phi_0)$ (or a negligible $A$-term discussed in
\cite{Jokinen1}), or a pocket of false minimum \cite{AEJM}. Neither
case leads to a slow roll inflation within the MSSM.}

For $\phi < \phi_0$ the mass term dominates the flat direction
potential.  Then quantum fluctuations can push $\phi$ to the vicinity
of $\phi_0$ if $\phi_{r.m.s} \geq \phi_0$.\footnote{Due to the
Gaussian distribution of fluctuations, the probability of having $\phi
\gg \phi_{r.m.s}$ is exponentially suppressed.} This, according to
(\ref{rms}), requires that
\be \label{cond1}
H_{\mathrm{false}} \geq \left(\frac{8 \pi^2}{3}\right)^{1/4} (m_{\phi}
\phi_0)^{1/2} \simeq 10^{9}~\mathrm{GeV}.
\ee
The number of e-foldings needed for this to happen is
\be \label{efold}
\N_{\mathrm{false}} \leq \left(\frac{H_{\mathrm{false}}}{10^9~\mathrm{GeV}}
\right)^2 10^{12}.
\ee
Indeed for $H_{\mathrm{false}} \geq \phi_0$ the inherent uncertainty due to
quantum fluctuations implies that $\phi > \phi_0$ within one Hubble time.

A last stage of MSSM inflation with an expansion rate
\be \label{hinf}
H_{\mathrm{MSSM}} =\frac{n-2}{\sqrt{6n(n-1)}}\frac{m_{\phi}\phi_{0}}
{M_{P}} \sim {\cal O}(1~{\mathrm{GeV}})
\ee
starts if $V(\phi)$ dominates the energy density of the universe,
i.e. $V(\phi) > 3 H^2_{\mathrm{false}} M^2_{P}$. An observationally
consistent inflation in the slow roll regime requires that the
displacement from the saddle point satisfy $| \phi - \phi_0 | < \Delta
\phi$, where~\cite{AEGM,AEGJM}
\be \label{domain}
\Delta \phi = \frac{\phi^3_0}{4n(n-1) M^2_{P}} \simeq
10^6~\mathrm{GeV}.
\ee
In the landscape, the universe can begin in a false vacuum with
arbitrarily large $H_{\mathrm{false}}$ (as long as $H_{\mathrm{false}}
\ll M_{P}$). Therefore, we generically expect that $\phi$ is quickly
pushed to field values $\phi \gg \phi_0$.  However,
$H_{\mathrm{false}}$ slowly decreases as a result of tunnelling to
vacua with smaller cosmological constant~\footnote{Note that we need
to stay in an MSSM-like vacuum all the way until MSSM inflation
begins. Given the scarcity of MSSM-like vacua in the landscape, the
probability of tunnelling from one such vacuum to another is $\sim
10^{-9}$.
However, due to eternal inflation in the false vacuum, the physical
volume of the universe increases by a factor of ${\rm exp}(3 H_{\rm
false}/\Gamma)$ within a typical time scale for bubble nucleation
($\Gamma$ is the false vacuum decay rate, see subsection 2.2). This
easily wins over the suppression factor $10^{-9}$ for $\Gamma < 9
H_{\rm false}$.}. For a massive scalar field with the potential
$V(\phi)\sim m^2_{\phi} \phi^2/2$, this implies a gradual decrease of
$\phi_{r.m.s}$, see (\ref{rms}). Since quantum fluctuations can at
most push $\phi$ to $\phi_{r.m.s}$, this also implies that $\phi$ is
slowly decreasing in time.  Indeed for $H_{\mathrm{false}} < 10^9$
GeV, we find that $\phi < \phi_0$, irrespective of how large $\phi$
initially was.  This is the case in the discussion of
\cite{hep-th/0004134}, which is why \cite{hep-th/0004134} requires a
jump from large cosmological constant directly to the slow-roll
inflationary stage.

Note, however, that the potential becomes very flat around $\phi_0$ as
a result of the interplay among different terms in
(\ref{hubblepot}). In fact, for $| \phi - \phi_0 | \ll \phi_0$ we have
\cite{AEGM,AEGJM}
\be \label{3rd}
V(\phi) \approx V(\phi_0) + \frac{1}{3(n-2)^2}
\frac{m^2_{\phi}}{\phi_0} (\phi - \phi_0)^3\,.
\ee
Once $H_{\mathrm{false}} \sim 10^9$ GeV, $\phi$ reaches this plateau
(from above).  Within the plateau, $V^{\prime}(\phi)$ becomes
increasingly negligible, and so does the classical slow roll; they
exactly vanish at $\phi = \phi_0$. In consequence, quantum jumps
dominate the dynamics and freely move $\phi$ throughout the
plateau. Hence the flatness of potential, which is required for a
successful MSSM inflation, guarantees that $\phi$ will remain within
the plateau during the landscape evolutionary phase.  It is possible
to generalize this argument to other models of low-scale inflation; we
see that a sufficiently flat inflaton potential can trap inflaton
fluctuations in the slow-roll region.

The flat direction eventually dominates the energy density of the
universe when $H_{\mathrm{false}} < 1$ GeV, see (\ref{hinf}). MSSM
inflation then starts in those parts of the universe which obey
(\ref{domain}). Having $\phi$ so close to $\phi_0$ is possible since
the uncertainty due to quantum fluctuations is $\sim 1$ GeV at this
time.

The false vacuum inflation paves the way for an MSSM inflation inside
the nucleated bubble.  The modulus which was responsible for the false
vacuum inflation continues tunnelling to minima with smaller
(eventually the currently observed) cosmological constant inside the
MSSM inflating bubble. The modulus will oscillate around the minimum
of its potential as the curvature of its potential dominates over the
Hubble expansion rate. The fate of oscillations would depend on the
coupling of the modulus to the MSSM fields. For a coupling of
gravitational strength, oscillations are long-lived. Moreover, the
decay will be kinematically forbidden if the decay products are
coupled to the flat direction $\phi$ (which has obtained a large VEV
$\simeq \phi_0$). However, such details are largely irrelevant as the
flat direction dominates the energy density of the universe and drives
inflation, diluting the energy density in oscillations.

In particular, we note that MSSM inflation has a self-reproducing
regime \cite{AEGM,AEGJM} because $V^{\prime}(\phi)$ is extremely small
and the potential becomes very flat close to $\phi_0$.  The observable
part of our universe can spend an arbitrarily long time in the
self-reproducting regime before moving into the standard slow-roll
inflation.  During this period, the cosmological constant can continue
to decay and eventually settle at an observationally acceptable value.

Before we conclude this discussion, let us remind the readers that
obtaining the flat direction VEV, $\phi_0$, depends on the false
vacuum inflation, which requires $H_{\mathrm{false}}\geq 10^{9}$ GeV
at some time (see (\ref{cond1})).  This condition, although very
probable in the landscape picture, need not be satisfied always. Then
the quantum fluctuations would not be large enough to push the flat
direction to the vicinity of $\phi_0$ as required for a final stage of
MSSM inflation. This would therefore ruin the inflationary and
phenomenological predictions. This can be avoided if the flat
direction is trapped in a false minimum which evolves with time, as
discussed below.


\subsubsection{Negative Hubble induced corrections}

The case with $c \sim -\mathcal{O}(1)$ may arise naturally for
non-minimal K\"ahler potential~\cite{dine95,dine96}. For
$H_{\mathrm{false}} \gg m_{\phi}$ the potential in (\ref{hubblepot})
becomes tachyonic, and its true minimum is located at
\be \label{true}
\phi_{\mathrm{min}} \simeq \left(\frac{\sqrt{| c |}}{\lambda_n \sqrt{2n-2}}
H_{\mathrm{false}} M_{P}^{n-3}\right)^{1/(n-2)}\,.
\ee
Note that $\phi_{\mathrm{min}}$ is initially larger than $\phi_0$, see
(\ref{saddle}).

The curvature of the potential at the minimum is $V^{\prime \prime}
(\phi_{\mathrm{min}}) = (4n-7) | c | H^2_{\mathrm{false}} \gg
H^2_{\mathrm{false}}$.  This implies that $\phi$ is driven away from
the origin, due to quantum fluctuations, and quickly settles down at
$\phi_{\mathrm{min}}$.  The MSSM flat direction is \textit{trapped}
inside the minimum, held due to false vacuum inflation, and gradually
tracks the instantaneous value of $\phi_{\mathrm{min}}$ as
$H_{\mathrm{false}}$ decreases.\footnote{Here we are assuming that the
difference between the energy densities of the two false vacua is
small compared to their average energy density.}

The minima at $\phi = 0$ and $\phi = \phi_{\mathrm{min}}$ become
degenerate when $H_{\mathrm{false}} \sim m_{\phi}$. For
$H_{\mathrm{false}} \ll m_{\phi}$ the true minimum is at $\phi = 0$
and the one at $\phi_{\mathrm{min}}$ will be \textit{false}. The
Hubble induced corrections are subdominant in this case. We then find
\cite{AEGJM}
\begin{eqnarray}
& & \phi_{\mathrm{min}} \simeq \phi_0 \left(1 + \frac{1}{n-2}
\frac{\sqrt{| c |} H_{\mathrm{false}}}{m_{\phi}}\right)
\, \\ \label{falmin}
& & \, \nonumber \\
& & V^{\prime \prime}(\phi_{\mathrm{min}}) \simeq 2 (n-2)
\sqrt{| c |} H_{\mathrm{false}} m_\phi \, . \label{falcurv}
\end{eqnarray}
The $\phi$ field can track down the instantaneous value of
$\phi_{\mathrm{min}}$ provided that $\sqrt{V^{\prime
\prime}(\phi_{\mathrm{min}})}$ is greater than the Hubble expansion
rate. In fact this is the case so long as $H_{\mathrm{false}} >
H_{\mathrm{MSSM}} \sim 1$ GeV, see (\ref{falcurv}). Once
$H_{\mathrm{false}} \simeq H_{\mathrm{MSSM}}$, the flat direction
potential dominates the energy density of the universe, and MSSM
inflation begins at a Hubble expansion rate $H_{\mathrm{MSSM}}$.  In
the meantime, landscape tunnelling to vacua with smaller cosmological
constant continues, and the location of the false minimum
$\phi_{\mathrm{min}}$ continuously changes.\footnote{Note that
tunnellings do not affect the Hubble expansion rate anymore since the
flat direction dominates the energy density of the universe now.}
Eventually $V^{\prime \prime}(\phi_{\mathrm{min}}) <
H^2_{\mathrm{MSSM}}$ when
\be \label{end}
H_{\mathrm{false}} \simeq \frac{1}{2(n-2)}
\frac{H^2_{\mathrm{MSSM}}}{m_{\phi}},
\ee
at which time
\be \label{falminend}
\phi_{\mathrm{min}} - \phi_0 \simeq \frac{\phi_0}{2(n-2)^2} \left(
\frac{H_{\mathrm{MSSM}}}{m_{\phi}}\right)^2.
\ee
It turns out from (\ref{hinf},\ref{domain}) that $\phi_{\mathrm{min}}
- \phi_0 \ll \Delta \phi$.  Therefore $\phi$ is already inside the
interval required for a successful MSSM inflation. At this point the
Hubble induced corrections become largely unimportant, and all the
successes of MSSM inflation are retained, as discussed in the previous
subsection. The fate of the string moduli inside the MSSM bubble
remains the same as in the previous subsection.  In particular, it
does not matter whether the universe tunnels right away to the
currently observed value of $\Lambda$ or not. Inflation dilutes any
excitations of the modulus oscillations during inflation.  Our Hubble
patch reheats when the MSSM flat direction rolls down to its minimum
and starts creating MSSM quanta as discussed in the previous
subsection.

Note that all needed for the success of this scenario is to start in a
false vacuum in the landscape where $H_{\mathrm{false}} > m_{\phi}
\sim 1$ TeV. This ensures that the flat direction will roll way from
the origin and settle at $\phi_{\mathrm{min}}$, which is the true
minimum of $V(\phi)$ at that time. It will then track
$\phi_{\mathrm{min}}$ as $H_{\mathrm{false}}$ slowly decreases, and
will eventually land inside the appropriate interval around $\phi_0$.
This is a much milder condition than that in the case of negligible
Hubble induced corrections $H_{\mathrm{false}} \geq 10^9$ GeV (see the
previous subsection).

One comment is in order. This scenario has some similarities to the
Affleck-Dine baryogenesis with negative Hubble induced corrections
\cite{dine95,dine96}. It can be seen from (\ref{hubblepot}) that the
equation of motion of the flat direction has a fixed point so long as
Hubble induced corrections are dominant. The flat direction tracks the
fixed point in a radiation or matter dominated universe.  However, a
non-adiabatic change in the potential occurs when $H \sim m_{\phi}$,
and the flat direction starts oscillating around the origin.  However,
in our case the universe is in a de Sitter phase with a slowly varying
$H_{\mathrm{false}}$, and the flat direction tracks the false minimum
of its potential until it dominates the universe. As explained, this
is necessary for having a successful MSSM inflation.

\section{Acknowledgments}

The author would like to thank all his collaborators: Rouzbeh
Allahverdi, Bhaskar Dutta, Kari Enqvist, Andrew Frey, Juan
Garcia-Bellido, Asko Jokinen and Alex Kusenko for successful
collaborations, invigorating, delightful and extremely helpful
discussions on various aspects of the physics involved in
understanding inflation and its consequences.

We also wish to thank Cliff Burgess, Manuel Drees, John Ellis, Jaume
Garriga, and Tony Riotto for valuable discussions and various
suggestions they have made. We also benefited from the discussions
with Shanta de Alwis, Steve Abel, Mar Bastero-Gil, Micha Berkhooz,
Zurab Berezhiani, Robert Brandenberger, Ramy Brustein, Kostas
Dimopoulos, Damien Easson, Renata Kallosh, Gordy Kane, Justin Khoury,
George Lazarides, Andrei Linde, Andrew Liddle, David Lyth, Hans Peter
Nilles, Pavel Naselsky, Lyman Page, Maxim Pospelov, Subir Sarkar,
Qaisar Shafi, Misha Shaposhnikov, Paul Steinhardt, Scott Thomas and
Igor Tkachev.

The research of A.M. is partly supported by the European Union through
Marie Curie Research and Training Network ``UNIVERSENET''
(MRTN-CT-2006-035863).


\end{document}